\pdfoutput=1
\documentclass[11pt]{article}

\usepackage[final]{EMNLP2023}

\usepackage{amsfonts} 

\usepackage{times}
\usepackage{latexsym}
\usepackage{multirow}
\usepackage{csquotes}
\usepackage{booktabs}
\usepackage{tikz}
\usepackage{pgfplots}

\renewcommand{\paragraph}[1]{\noindent\textbf{#1}}

\usepackage[T1]{fontenc}
\usepackage[utf8]{inputenc}

\usepackage{microtype}

\usepackage{inconsolata}

\title{How Does Generative Retrieval Scale to Millions of Passages?}
\author{Ronak Pradeep\thanks{\ \ Equal Contribution.}\ \ \thanks{\ \ Work completed while a Student Researcher at Google.}\ \ $^\mathsection$, Kai Hui$^*$, Jai Gupta, Adam D. Lelkes, Honglei Zhuang\\
\textbf{Jimmy Lin$^\mathsection$, Donald Metzler, Vinh Q. Tran$^*$} \\
  Google Research, $^\mathsection$University of Waterloo\\
  \texttt{rpradeep@uwaterloo.ca},\ \texttt{\{kaihuibj,vqtran\}@google.com} \\}

\begin{document}
\maketitle

% \author{Anonymous Author(s)}

% \author{Kai Hui}
% \email{kaihuibj@google.com}
% \affiliation{%
%   \institution{Google Research}
%   \city{Berlin}
%   \country{Germany}
% }

% \renewcommand{\shortauthors}{X. Hui et al.}

\begin{abstract}
Popularized by the Differentiable Search Index, the emerging paradigm of generative retrieval re-frames the classic information retrieval problem into a sequence-to-sequence modeling task, forgoing external indices and encoding an entire document corpus within a single Transformer.
Although many different approaches have been proposed to improve the effectiveness of generative retrieval, they have only been evaluated on document corpora on the order of 100k in size.
We conduct the first empirical study of generative retrieval techniques across various corpus scales, ultimately scaling up to the entire MS MARCO passage ranking task with a corpus of 8.8M passages and evaluating model sizes up to 11B parameters.
We uncover several findings about scaling generative retrieval to millions of passages; notably, the central importance of using synthetic queries as document representations during indexing, the ineffectiveness of existing proposed architecture modifications when accounting for compute cost, and the limits of naively scaling model parameters with respect to retrieval performance.
While we find that generative retrieval is competitive with state-of-the-art dual encoders on small corpora, scaling to millions of passages remains an important and unsolved challenge.
We believe these findings will be valuable for the community to clarify the current state of generative retrieval, highlight the unique challenges, and inspire new research directions.

\end{abstract}

\section{Introduction}\label{sec.intro}
For the last several years, dual encoders~\citep{gillick2018end,karpukhin2020dense,ni22gtr,chen2022towards} have dominated the landscape for first-stage information retrieval.
They model relevance by mapping queries and documents into the same embedding space, optimized via contrastive learning~\citep{Hadsell2006DimensionalityRB, gao-etal-2021-simcse}. 
Dense embeddings are pre-computed for all documents in a corpus and stored in an external index. 
This allows for fast approximate nearest neighbor search~\citep{Vanderkam2013NearestNS, JonhsonKNN} to retrieve relevant documents. 
Cross-encoders based on large Transformer models~\citep{monobert, nogueira-etal-2020-document, Pradeep2021TheED} often function on top of these retrieved documents to further refine the top results.

Recently, the emerging paradigm of generative retrieval~\citep{cao-aer, Tay2022TransformerMA} sought to replace this entire process with a single sequence-to-sequence Transformer model~\citep{sutskever2014sequence, vaswani2017attention}, showing promising results against dual encoders given a sufficiently small corpus size.
Since then, various techniques, such as~\citep{zhuang2022bridging,autoregressive-search-engine,ultron,Wang2022ANC,chen2023understanding}, have aimed to improve the effectiveness of generative retrieval models, either with alternative document identifier formulations, architecture changes, or training objectives.
Such work, however, has only evaluated generative retrieval over relatively small corpora on the order of 100k documents, such as Natural Questions~\citep{kwiatkowski2019natural}, TriviaQA~\citep{joshi2017triviaqa}, 
% KILT \citep{petroni-etal-2021-kilt}, 
or small subsets of the MS MARCO document ranking task~\citep{nguyen2016ms}. 
Despite these research contributions, a number of open questions remain unanswered, including \textit{how well} current generative retrieval techniques work on larger corpora and \textit{which aspects} of generative retrieval models proposed so far are vital at scale.

In this paper, we conduct the first empirical study of generative retrieval techniques over the entire MS MARCO passage-level corpus, evaluating its effectiveness over 8.8M passages.
We select popular approaches in recent works and evaluate them first on Natural Questions and TriviaQA to establish a definitive ablation of techniques in a controlled setup. 
Our experiments mainly focus on evaluating techniques proposed by \citet{Tay2022TransformerMA}, \citet{zhuang2022bridging}, and \citet{Wang2022ANC}.
Namely, we ablate document identifier design: atomic, naive, semantic; document representation design: document tokens, ground truth queries, synthetic queries \citep{nogueira2019doc2query}; and model design: prefix-aware weight-adaptive decoding, constrained decoding, and consistency loss during training. At this small scale, we demonstrate state-of-the-art results for retrieval, generative and non-generative, over the NQ variant from \citep{Wang2022ANC}, without the need for many proposed methods. 

We then scale up the corpus size leveraging the MS MARCO passage ranking task, beginning with a subset of 100k passages before increasing the count to 1M and 8.8M passages (the entire set).
Incrementally doing so allows us to establish which techniques remain effective as corpus size and difficulty scale. 
Finally, to explore the effect of model scaling on retrieval effectiveness on large corpora, we select a set of techniques with promising results at T5.1.1-Base scale \citep{raffel2020exploring} and modify the parameterization to consider up to 11B parameters.
As the parameter distributions vary between methods, e.g., Atomic IDs cost embedding dimension times corpus size parameters, while Naive IDs do not cost anything beyond the core Transformer model, we aim to provide some insight into the trade-off of different parameter allocations on a large corpus. 

While our experimental findings are nuanced, we summarize the main findings as follows:
\begin{enumerate}
    \item Of the methods considered, we find synthetic query generation to be the single most critical component as corpus size grows. 
Defining the task of generative retrieval as solely mapping from synthetic queries to document identifiers is the most effective modeling strategy, with all other modeling strategies largely unnecessary.
    \item As corpus size increases, discussion of compute cost is crucial.
Methods that implicitly increase model parameters perform better using the same T5 initialization. 
However, the quality improvements vanish as we scale up the naive approach to similar parameter sizes. 
Following~\citep{dehghani2022the}, we note that the parameter count is not the entire story and provide more discussion regarding model comparisons and trade-offs in Section~\ref{sec.discussion-scaling}.
    \item Increasing the model size is necessary for improved generative retrieval effectiveness.
However, somewhat surprisingly, for the best sequential IDs, effectiveness does not improve past a certain point -- peaking at XL (3B) with a slightly worse score using XXL (11B) under fixed experimental settings. 
We find this counter-intuitive to the common conception of generative retrieval being limited by model capacity.
\end{enumerate}

\noindent Our findings conclude that on the entire MS MARCO passage ranking task, simply scaling a model trained solely on synthetic queries to Naive ID generation demonstrates the best effectiveness of all techniques considered. 
On a small subset of 100k passages, a T5-Base model trained with this strategy achieves 82.4 MRR@10 (Section \ref{sec.discussion-queries}), competitive with GTR-Base \citep{ni22gtr} at 83.2 MRR@10. 
While on the 8.8M passages, a T5-XL model trained with this approach achieves only 26.7 MRR@10.

While the field of generative retrieval continues to evolve rapidly, it is clear that achieving competitive effectiveness against state-of-the-art dense retrieval models at scale remains an important and unsolved challenge.
Our results suggest the need for continued research into generative retrieval and more fundamental advances to the paradigm before we are able to fully leverage the power of scaling up model parameters.
We believe that our findings will help the research community better understand the current challenges faced when applying generative retrieval models to larger corpora and inspire new research in this direction.

\section{Related Work}
\label{sec.related_work}
Traditional retrieval models like BM25~\citep{robertson2009probabilistic} that rely on the lexical overlap, term frequency heuristics, and inverse document frequency, while reasonably strong on their own, tend to fail at matching documents that have minor word overlap but are semantically related.

A popular solution is dual encoders \citep{gillick2018end, karpukhin2020dense, chen2022towards}, where a pretrained language model such as BERT~\citep{devlin2019bert}  is used to compute low-dimensional dense representations instead of the high-dimensional sparse representations found in BM25.
These dual encoder models are further trained on the target task to achieve improved effectiveness.
Based on the success of T5 in various natural language understanding tasks, \citet{ni22st5} proposes scaling up dual encoders by training T5-style pretrained language models with a two-stage contrastive learning approach on the Semantic Text Similarity (STS) tasks.
The Generalizable T5 Retriever (GTR)~\citep{ni22gtr} extends this idea to information retrieval.
The most successful GTR models were pretrained on a large-scale question-answering dataset curated from the internet and fine-tuned on the MS MARCO Passage Ranking task~\citep{nguyen2016ms}.

Existing approaches often apply synthetic query generation to improve retrieval effectiveness.
\citet{nogueira2019doc2query} first leveraged a vanilla sequence-to-sequence Transformer to train a model that can map passages to queries that it might be able to answer. 
\citet{nogueira2019docT5query}, doc2query-T5 further improved the effectiveness of the traditional Transformer by leveraging a T5 model. \citet{Ma2022DocumentEB} experimented with similar ideas and showed that query generation is effective across a wide range of corpora and task setups.
%In more recent work,~\citet{d2q-less-more} examined if the effectiveness of doc2query-T5 can be further improved by filtering out queries deemed to be ``hallucinated'' content based on the scores assigned by a powerful pointwise ranker to the query and passage pair.
%They found that such filtering can indeed help improve the retrieval effectiveness significantly while simultaneously reducing the mean execution time and the index size.
%Since neural retrieval models are known to require massive supervised training sets to surpass traditional term-based methods, \citet{ma-etal-2021-zero} proposed a zero-shot technique that uses synthetic question generation to bridge the gap and further improve effectiveness with a hybrid dense-sparse retrieval.

Prior to generative retrieval, sequence-to-sequence language models, like T5~\citep{2020t5}, were shown to be effective for reranking tasks.
In this setup, models assign scores to the top-$k$ results from a first-stage retrieval method. 
One can then use these scores to rerank the documents. 
For example, monoT5~\citep{nogueira-etal-2020-document} was the first to leverage T5 as a pointwise reranker by training a model that takes the concatenation of the query and document as input and generates a relevance label. 
\citet{Pradeep2021TheED, zhuang22rankt5, hui-etal-2022-ed2lm} have since improved the performance and efficiency of generation-based reranking.
These approaches continue to demonstrate strong effectiveness~\citep{craswell2022overview,pradeep21vera, pradeep22ct}.

Generative retrieval seeks to replace the entire information retrieval process with a single sequence-to-sequence model capable of mapping queries directly to relevant document identifiers~\citep{metzler21rethinking}.
Differentiable Search Indexes (DSI)~\citep{Tay2022TransformerMA} first demonstrated the potential of this paradigm, where T5 is used to parameterize an end-to-end search system, with the model parameters encoding all information about the corpus.
See Section~\ref{sec.methods} for more information.
DSI was shown to outperform a dual encoder baseline on Natural Questions dataset~\citep{kwiatkowski2019natural}.
\citet{zhuang2022bridging} explores the effectiveness of DSI and synthetic queries on a 100k passage subset of the MS MARCO passage ranking corpus and XOR QA~\citep{asai-etal-2021-xor}.
Neural Corpus Indexer~\citep{Wang2022ANC} builds on the success of DSI and introduces a combination of more input variants and architectural additions, some of which we describe and explore in this work. 
Many works have explored various document identifier designs, including document substring~\citep{autoregressive-search-engine}, metadata-based approaches~\citep{ultron, ziems2023large}, and learned quantization~\citep{dsi-recsys, sun2023learning}. 
More recently,~\citep{chen2023understanding} proposes a distillation approach on top of DSI, learning from the rankings generated by dense retrieval using a multi-task training loss.

However, none of these works have explored training or evaluating generative retrieval systems on corpora larger than O(100k) documents. 
Given that the generative retrieval paradigm has extended beyond traditional information retrieval into areas such as recommender systems~\citep{dsi-recsys} and vision~\citep{zhang2023irgen}, we believe our study on scaling will be crucial for an evergrowing community.

\section{Methods}\label{sec.methods}
In this section, we revisit the design details of the generative information retrieval method, using the Differentiable Search Index (DSI)~\citep{Tay2022TransformerMA} as the baseline.
Then, we describe multiple techniques introduced in subsequent works that we aim to ablate and study in this work~\citep{Wang2022ANC, zhuang2022bridging}.
% These essential design choices and components
% are summarised in Table~\ref{tab.method}.

\subsection{Background}\label{sec.backbone}
% Different from dual encoder~\citep{karpukhin2020dense} or cross-attention re-ranker~\citep{nogueira2020T5ranking},
DSI~\citep{Tay2022TransformerMA} reformulates the retrieval task as a sequence-to-sequence (seq2seq) task, with queries as inputs and document identifiers (docids) relevant to the query as generation targets.
The corpus, namely the mapping between the document's content and its identifier, is encoded using the parameters of the LLM. 
DSI achieves this by leveraging two seq2seq tasks: indexing and retrieval.
During training, the model learns to generate the docid given the document content (indexing task) or a relevant query (retrieval task).
%The most natural way to formulate the retrieval task is to use existing labeled data for training. We denote this as "Labels" in our experiment tables.
At inference time, the model processes a query and generates a ranked list of identifiers as retrieval results.

\subsection{Inputs and Targets}\label{sec.inputstargets}
In the framework discussed, DSI learns to encode the mapping between the long-form textual representation of a document
and its identifier in its parameters while also learning to fetch the same identifier when it receives a relevant query as input.

Two crucial design choices are how documents are represented (i.e., the inputs in the indexing task) and how document identifiers (docids) are represented (i.e., the targets in both indexing and retrieval tasks).
Two primary considerations are: (1) For document representations, it is prohibitive to encode long textual sequences with a Transformer~\citep{vaswani2017attention}-based LLM, making it difficult to index full documents and
(2) The naive identifiers taken from an existing dataset could be sub-optimal, for instance, due to their lack of semantic meaning.
In this work, we consider different design choices for both these components.

\subsubsection{Document Representations}
One straightforward idea is to pick a text span from the document as a representation. 
DSI considers the first 64 tokens (FirstP) in each document, whereas \citet{Wang2022ANC} leverages ten randomly-selected chunks of 64 consecutive tokens, a technique they call Document As Query (DaQ).
When working with Natural Questions and TriviaQA, which contain lengthy documents, we examine each variant separately and in combination.
In the case of MS MARCO, which has short passages, FirstP and DaQ are essentially the same, assuming sufficient context length.

\subsubsection{Synthetic Query Generation}
For training the model for the retrieval task, the natural baseline uses existing labeled data, i.e., queries from the retrieval dataset as inputs and the docids labeled as relevant as targets (we will denote this as "Labeled Queries" in our tables).

However, as argued in~\citet{zhuang2022bridging} and~\citet{Wang2022ANC},
there are two kinds of gaps between the index and retrieval tasks. First is the data distribution gap: queries for the retrieval task are short and request specific information, while the documents for the indexing task are long and convey information.
Second is the coverage gap: the model is exposed to the entire corpus during the training of the indexing task, while only positive examples have associated queries in the retrieval task.
The latter problem is exacerbated in the MS MARCO passage ranking task as only 550K passages have an associated query for training the retrieval task, while the indexing task has to learn to encode all 8.8M passages in the corpus.

Their proposed method for mitigating this gap is by generating synthetic queries for each document using a query generation model such as docT5query~\citep{nogueira2019docT5query}.
The generative retrieval model is then trained to predict the docid given the corresponding synthetic queries.
We can also think of these synthetic queries as alternative document representations.

\subsubsection{Document Identifiers}
In this work, we consider four kinds of different identifiers: the three kinds of document identifiers from the original DSI paper: unstructured atomic identifiers (Atomic IDs), naive string identifiers (Naive IDs), and semantically structured identifiers (Semantic IDs), and the 2D Semantic IDs from \citet{Wang2022ANC}.
\\\\
\noindent\textbf{Atomic IDs.} We treat each docid as a single, or ``atomic'' token in this setting. 
The decoder, then, only needs to run for a single decoding step; we then sort the logits of the docids to obtain the ranked document list.
The setting requires adding a token, for each document, to the model vocabulary, increasing the model's parameter count by corpus size times embedding dimension, which can be expensive for large corpora.
When considering millions of documents, we apply two optimizations to make implementation more feasible. 
First, the encoder's embedding table is adjusted to only consist of the standard T5 vocabulary, while the decoder's output projection only corresponds to docids. 
Second, we take special care to ensure the output projection is properly sharded across cores to distribute memory cost to allow scaling.
In the \texttt{t5x}  framework~\cite{roberts2022t5x}, this corresponds to setting appropriate partitioning rules.
\\\\
\noindent\textbf{Naive IDs.} In this setting, the original document identifier from a corpus is directly used and treated as a textual string.
For example, a five-digit number ``42915'' is treated as a string and passed through the SentencePiece vocabulary of T5.
It is worth noting that such naive document identifiers might also capture some semantics of the corpus, as they depend on the curation pipeline that might leak some notions of relatedness.
\\\\    
\noindent\textbf{Semantic IDs.} 
Following \citet{Tay2022TransformerMA}, instead of relying on predefined identifiers, Semantic IDs aim to imbue document identifiers with hierarchical semantic information.
Specifically, after encoding documents into dense vectors, a hierarchical $k$-means algorithm recursively clusters the space into $k$ clusters until individual clusters include no more than $c$ documents. 
Consequently, all document identifiers form a tree, where non-leaf nodes correspond to super-clusters, and leaf nodes are clusters with at most $c$ documents each.
Semantic IDs are formed by composing these cluster ids, each from $0$ to $k-1$, tailed by a document id in the leaf nodes between $0$ and $c-1$. 
In this work, we use the identifiers generated by \citet{Wang2022ANC} for NQ and TriviaQA for a fair comparison. 
These are based on a 12-layer BERT model.
For MS MARCO, we use SentenceT5-Base \cite{ni22st5}, and $c=100$. 
Since the passage-level corpus is large, if a cluster ends up bigger than 1M documents, we sample 100k when computing centroids. 
We used $k=10$ clusters at each level, corresponding to the ten digits ($0\ldots9$).
\\\\
\noindent\textbf{2D Semantic IDs.} In the Semantic ID setting, the same tokens are used to represent different semantic meanings at different positions: we use the same set of numbers/tokens between $0$ to $k-1$ for all identifiers, but they represent semantic clusters at different levels in the tree.
To address this, NCI~\citep{Wang2022ANC} extends the Semantic ID and introduces its 2D variant by adding an extra dimension to encode the positions, making the model aware of levels of clustering when decoding the identifier. 
To implement this modeling change, they additionally introduce a change to the decoder described in the next section.

% https://source.corp.google.com/piper///depot/google3/experimental/users/ronakpradeep/neural_corpus_indexer/nci_architecture.py;l=344#:~:text=343-,344,-345
\subsection{Model Variants}\label{sec.nci}
Besides alternative ways of constructing model inputs and targets, generative retrieval approaches that build on DSI have also investigated novel modeling components.  
Here, we review three model components introduced by \citet{autoregressive-search-engine} and \citet{Wang2022ANC}.
% They are designed based on 2D Semantic ID.
% The results in NCI are mostly based on 2D semantic id, and so are the components described below.
\\
\\
\paragraph{Prefix-Aware Weight-Adaptive Decoder (PAWA)}
is proposed as a method for decoding 2D Semantic IDs.
Unlike a standard Transformer decoder, which uses the same matrix to project the decoder's hidden representation to the vocabulary space for every position, PAWA uses different projection matrices at each timestep, with the weights of each projection matrix computed adaptively by a separate Transformer decoder.
%In a nutshell, when projecting the dense representation from the last multi-head attention block to the vocabulary space for embedding lookup, instead of sharing a projection matrix PAWA uses different projection matrices at different decoding positions, so the name ``adaptive''.
Specifically, in a vanilla decoder, the dense representation $h\in \mathbb{R}^{l \times d}$ from the last decoder layer is projected into the vocabulary space with $W\in \mathbb{R}^{d\times |V|}$, where $l$ denotes the sequence length for decoding.
To incorporate the position, the extra decoder in PAWA separately processes the input query and the already-decoded docid tokens to output a projection matrix  $W^{pawa}\in\mathbb{R}^{d \times l \times |V|}$, replacing $W$.
This aims to capture that the semantic meaning of a docid token depends on its position in the output sequence as well as on the docid prefix preceding it.
%  the projection matrix $W$ with $W^{pawa}\in\mathbb{R}^{d \times l \times |V|}$.
 %When using $W^{pawa}$ to project $h$, each position in the sequence uses different projection  matrix $W^{pawa}_i \in \mathbb{R}^{d \times 1 \times |V|}$.
 The experiments in this paper use the open-source PAWA implementation provided by the original authors\footnote{\url{https://github.com/solidsea98/Neural-Corpus-Indexer-NCI}} as a reference and build it out on \texttt{t5x}.
For more details, one could refer to~\cite{Wang2022ANC} and their code base.
\\\\
\paragraph{Constrained decoding} can be used to avoid generating invalid document identifiers~\citep{autoregressive-search-engine, Wang2022ANC}. 
A potential reason is that the space of identifiers is sparse, especially for Semantic IDs, and constrained decoding may help with memorization.
While we have empirically found that roughly less than 1 in 20 DSI-based generation beams are invalid, we include this method nonetheless, as it is widespread in the literature.
In this work, we adopt an exact match approach that leverages a trie to ensure only valid document identifiers are decoded.
%We use a beam size of $100$ and a brevity penalty set to $0.0$ as docids of different lengths are equally preferred.
\\
\\
\paragraph{Consistency loss} can be used to alleviate over-fitting by 
introducing a regularization term.
The basic idea is that the representations generated by two forward passes with different dropout masks should be similar.
\citet{Wang2022ANC} incorporate this insight into a regularization term that augments the generation loss.
We investigate the softmax version as described in the NCI paper (Eq. 5 in~\citep{Wang2022ANC}) and a KL-divergence version from an early version of NCI (Eq.~\ref{eq.kl_consistency}).
They compute the Kullback-Leibler (KL) divergence between the output probabilities of two independent forward passes per position, where $p_{i,1}$ and $p_{i,2}$ are the probability distributions over the vocabulary space from the two forward passes at position $i$, respectively.
\begin{equation}
\resizebox{0.88\linewidth}{!}{
    $\mathcal{L}_{reg} =  \frac{1}{2} [D_{KL}(p_{i,1} \mid \mid p_{i,2}) + D_{KL}(p_{i,2} \mid \mid p_{i,1})]$
}
\label{eq.kl_consistency}
\end{equation}
While we closely follow the implementation of the Neural Corpus Indexer code base, we find that these regularization terms lead to training instability and that the model effectiveness often diverges into a $NaN$ loss. 
As a result, we do not include consistency regularization in our final experimental setup.

%\subsection{Putting It Together}\label{sec.method_summary}
%The described design choices and model components are summarised in Table~\ref{tab.method}.
%It can be seen that DSI and NCI share the backbone with the same index/generation tasks.
%DSI investigates the uses of Atomic, Naive, and, Semantic docid, whereas 
%NCI extends the Semantic docid into a two-dimensional docid.
%Different docid variants are exclusively used. 
%As for the document representations, each representation
%corresponds to one index task, mapping different document representations to the docid.
%In the bottom of Table~\ref{tab.method},
%three model components from NCI are listed, where PAWA and consistency loss are involved in training,
%whereas constrained decoding only affects inference. 
%The rightmost column describes a model that includes components contributing positively, coined as UGIR following the naming
%from~\citet{metzler21rethinking}.

\section{Experimental Setting}\label{sec.setting}
\begin{table}[t]
    \centering
    \resizebox{\linewidth}{!}{
    \begin{tabular}{c|cc}
    \toprule
         Dataset &  \#Docs & \vtop{\hbox{\strut \% Covered by}\hbox{\strut train query set}}\\
    \midrule
         NQ100k \cite{Wang2022ANC} & 110k & 98.4\%\\
         TriviaQA \cite{Wang2022ANC} & 74k & 57.7\%\\
         MSMarco100k & 100k & 92.9\%\\
         MSMarco1M & 1M & 51.6\%\\
         MSMarcoFULL & 8.8M & 5.8\%\\
    \bottomrule
    \end{tabular}}
      \caption{The coverage statistics of the benchmark datasets and their training query sets.}
    \label{tab.dataset}
\end{table}
We limit ourselves to English retrieval tasks, focusing on the behavior of generative retrieval models at varying corpus scales.

\subsection{Corpus and Training Data}\label{sec.corpus}
Following small-scale generative retrieval experiment setups~\cite{Tay2022TransformerMA, Wang2022ANC, zhuang2022bridging,chen2023understanding},
we start with experiments on the
Natural Questions~\cite{kwiatkowski2019natural} and TriviaQA~\cite{joshi2017triviaqa} datasets.
To better understand how different model configurations perform at scale and in more practical settings, we also experiment with variants of the MS MARCO passage ranking dataset.
The MS MARCO passage ranking dataset consists of a corpus of 8.8M passages and a training set of 532K queries. From this dataset, we construct three variants, namely,
MSMarco100k (100k passages), MSMarco1M (1M passages), and MSMarcoFull (all 8.8M passages).
It is worth noting that most documents in NQ100k and MSMarco100k have at least one relevant query in the training set. 
However, as we scale to MSMarcoFull, the fraction of documents with queries in the training set drastically drops to around $6\%$, leading to a more practical setup.
We summarize the statistics of these datasets in Table~\ref{tab.dataset}.
\\\\
\paragraph{NQ100k and TriviaQA.}
To enable comparisons, we reuse the documents, the segmented documents, the training/testing splits, and generated query sets from \citet{Wang2022ANC}.
% and the generated questions from their repository but generate our semantic identifiers as we diverge slightly in its creation methodology.
The Natural Questions and TriviaQA datasets have corpora of sizes $109$K and $74$K, respectively. 
Note that \citet{Wang2022ANC} refers to NQ100k as NQ320k; we refer to the number of \emph{unique} documents instead of the labeled training data size.
Most documents in the NQ100k dataset have at least one relevant question in the training data, while $58\%$ of the TriviaQA dataset has this property.
\\\\
% \begin{itemize}
\paragraph{MSMarco100k.} 
% Similar to~\citet{Zhuang2022BridgingTG}, 
In the same vein as NQ100k and TriviaQA, we curate a dataset with 100k passages sampled from the full MS MARCO passage ranking dataset.
Most passages have at least one positive query for training. 
We also include passages relevant to the queries in the development dataset (for evaluation).
\\\\
\paragraph{MSMarco1M.} This dataset is $10\times$ larger than MSMarco100k. 
As with MSMarco100k, we augment the corpus with passages relevant to development queries. 
We first include all passages relevant to the 533K and 7K queries from the training dataset and development sets, respectively.
This results in 516K and 7K unique passages from each set.
We randomly sample passages without a query in either set to total a million passages.
%As a result, this dataset setting diverges from all our smaller-scale datasets.
\\\\
\paragraph{MSMarcoFULL.} In this setting, we note another order of magnitude scale-up in corpus size.
As a result, only $5.8\%$ of the passages have a corresponding query in the training set.
We aren't aware of any previous work that has attempted to apply generative retrieval models to a dataset of this size and complexity.

\subsection{Synthetic Query Generation}\label{sec.qgen}
For NQ100k and TriviaQA, we reuse the generated questions from \cite{Wang2022ANC} with 20 and 15 generated questions for each document, respectively.
For the MSMarco variants, we use docT5query~\cite{nogueira2019docT5query} to generate questions, with 40 generated questions per passage. 
We also train a question-generation model using T5-base using training data from DPR~\cite{karpukhin2020dense}, a retrieval dataset derived from NQ~\cite{kwiatkowski2019natural}.
We use this model to generate 40 questions per passage, following the configuration of docT5query.
We refer to this variant as ``in-domain D2Q'' for NQ and TriviaQA.

\subsection{Evaluation Dataset and Metrics}\label{sec.metrics}
We report evaluation results on the development sets of each dataset.
For NQ100k and TriviaQA, the evaluation dataset includes 7830 and 7993 questions each.
For the three MSMarco variants, we use the validation split from the MS Marco passage ranking dataset, with 6,980 examples. For each query in the development sets, we use the models to generate ranked lists of documents. We report Recall@1 as the primary metric for Natural Questions and Recall@5 for TriviaQA.
For the MS MARCO passage ranking variants, we use Mean Reciprocal Rank at 10 (MRR@10) as our primary metric.

\subsection{Model Variants}\label{sec.model-settings}
We evaluate all methods using a T5.1.1 backbone~\citep{raffel2020exploring}. We test variants of labeled vs.\ synthetic queries, FirstP vs.\ DaQ document representations, as well as combinations of multiple representations. 
For each model variant, we ablate all versions of document identifiers when applicable. 
Model architecture additions are performed, in a stacking fashion, starting with the base model and then adding on PAWA, constrained decoding, and consistency loss in this order. 
Note, we only evaluate PAWA with 2D Semantic IDs, as it is built specifically for that setting.

For model scaling experiments, we mainly investigate whether Atomic IDs are an effective way to scale to millions of passages, given the parameter cost.
As such, we consider larger models with Naive IDs and Semantic IDs comparable to T5-Base with Atomic IDs, which total 7B parameters when scaling to 8.8M docids.

For baselines we provide BM25 \citep{robertson2009probabilistic} and BM25 with doc2query-T5 \citep{nogueira2019docT5query}.
For Natural Questions and TriviaQA, we also include the previous results reported for the NCI-variant of NQ (i.e., NQ100k). 
This includes state-of-the-art generative retrieval results like NCI and GenRet \citep{sun2023learning}, as well as GTR-Base, a state-of-the-art dual encoder \citep{ni22gtr}. 
For the new MS MARCO variants, we provide our own GTR-Base \cite{ni22gtr} results.

\begin{table*}[t!]
\centering
\resizebox{\textwidth}{!}{
\begin{tabular}{lllrrrrrr}
\toprule
& & &  \multicolumn{3}{c}{\textbf{NQ100k}} & \multicolumn{3}{c}{\textbf{TriviaQA}} \\
& \multicolumn{2}{l}{{\bf Model}} & At.\ & Nv.\ & Sm.\ & At.\ & Nv.\ & Sm.\\
\toprule
\multicolumn{3}{l}{\emph{Baselines}}\\
\multicolumn{2}{l}{BM25 (via \citet{Wang2022ANC})} & - & 15.1 &  - & - & 56.9 &  -   \\
\multicolumn{3}{l}{BM25 w/ doc2query--T5 (via \citet{Wang2022ANC}}  & - & 35.4 &  - & - & 59.7 &  -    \\
% \midrule
\multicolumn{3}{l}{{GTR-Base (via \citet{sun2023learning})}} & -  & 56.0 & - &  - & - & -   \\
\multicolumn{3}{l}{{NCI \citep{Wang2022ANC}}} & -  & 62.8 & 65.9 &  - & 88.8 & 90.5   \\
\multicolumn{3}{l}{{GenRet \citep{sun2023learning}}} & -  & - & 68.1 &  - & - & -   \\
%(b4) & \multicolumn{2}{l}{\emph{DSI\textsubscript{idx + gen} (reported)}} & -  & - & 27.4  &  - & - & - \\
% some Variants should follow but should they be reported / experiments we run \\
% (2b) & \multicolumn{2}{l}{w/o gen}  & - & - &  - & - & - &  -   \\
% (2c) & \multicolumn{2}{l}{w/o d2q + idx/daq}  & - & - &  - & - & - &  -    \\
% (2d) & \multicolumn{2}{l}{w/o reg}  & - & - &  - & - & - &  -    \\
% (2e) & \multicolumn{2}{l}{w/o PAWA}  & - & - &  - & - & - &  -    \\
% (2f) & \multicolumn{2}{l}{w/o constrained}  & - & - &  - & - & - &  -    \\
\midrule
\midrule
\multicolumn{3}{l}{\emph{Ours}}\\
(1a) & \multicolumn{2}{l}{Labeled Queries (No Indexing)} & 50.7 & 49.2 & 49.0 & 60.9 & 56.7 & 61.4\\ %https://tensorboard.corp.google.com/compare/nq_semantic_dual:4924499195590875337,trivia_gen:4551911958075541492,trivia_origin_gen:7680604033700664887,trivia_dual:7893897486184391027,nq_gen:6559200483537002674,nq_dual:1109478581885279694,trivia_origin_dual:1855148670792963707,nq_origin_gen:107723739900399712,nq_origin_dual:6634596749271406284,trivia_semantic_gen:1047534627324244553,trivia_semantic_dual:5740540757889352735,nq_semantic_gen:6348978493536113164/?darkMode=true&forceSVG=true&runFilter=nq#timeseries
(2a) & \multicolumn{2}{l}{FirstP + Labeled Queries (DSI)} & 60.0 & 58.4 & 58.7  & 71.6 & 75.2 & 78.9  \\
(2b) & \multicolumn{2}{l}{DaQ + Labeled Queries}  & 61.4 & 60.4 &  60.0 & 81.0 & 80.4 &  77.6    \\
(3a) & \multicolumn{2}{l}{DaQ + D2Q + Labeled Queries}  & 69.6 & 67.9 &  67.9 & 88.2 & 85.7 &  86.3    \\
(3b) &  \multicolumn{2}{l}{FirstP + DaQ + D2Q + Labeled Queries} & 69.0 &   68.2 & 67.2 & 88.9 & 86.9 & 87.4    \\
(4a) & \multicolumn{2}{l}{3b + PAWA (w/ 2D Semantic IDs)} & - & - &  66.3 & - & - &  86.5   \\
(4b) & \multicolumn{2}{l}{3b + Constrained Decoding} & - & - &  67.3 & - & - &  87.3   \\
(5) & \multicolumn{2}{l}{4b + Consistency Loss (NCI)} & - & - &  66.3 & - & - &  86.6 \\
\midrule
(6a) & \multicolumn{2}{l}{DaQ Only}  & 17.1 & 18.4 &  15.6 & 41.0 & 31.3 &  20.6   \\
% (3b) & \multicolumn{2}{l}{NCI\textsubscript{gen}}  & - & - &  - & - & - &  -    \\
% (2b) & \multicolumn{2}{l}{NCI\textsubscript{daq + gen}}  & 61.4 & 60.4 &  60.0 & 81.0 & 80.4 &  77.6    \\
% (2c) & \multicolumn{2}{l}{NCI\textsubscript{daq + gen + d2q}}  & 69.6 & 67.9 &  67.9 & 88.2 & 85.7 &  86.3    \\
% (2d) & \multicolumn{2}{l}{NCI\textsubscript{daq + gen + d2q + idx}}  & 69.0 &   68.2 & 67.2 & 88.8 & 86.9 & 87.0    \\
% \midrule
(6b) & \multicolumn{2}{l}{D2Q Only} & 43.6 & 42.3 & 42.9 & 61.9 & 57.8 & 57.1 \\
% \midrule
% (4a) & \multicolumn{2}{l}{\ \ + PAWA}  & - & - &  66.3 & - & - &  87.7   \\
% (4b) & \multicolumn{2}{l}{\ \ + Constrained Decoding}  & - & - &   & - & - &  87.0   \\
% \midrule
(6c) & \multicolumn{2}{l}{6b + PAWA (w/ 2D Semantic IDs) + Constrained Decoding} & - & - & 43.1 & - & - & 57.7 \\
\midrule
% \midrule
% (5) & Closest NCI (+ PAWA + ConDec)\\
% (3c) & \multicolumn{2}{l}{\ \ \ \ \ \ + Consistency Loss (Or NCI)} & - & - &  - & - & - &  -   \\
(7) & \multicolumn{2}{l}{3b + in-domain D2Q} & \textbf{70.7} & \textbf{69.7} & \textbf{69.5} & \textbf{90.0} & \textbf{88.0} & \textbf{89.2}\\
% (3i) & \multicolumn{2}{l}{DSI\textsubscript{d2q+daq} (PAWA + constrained)}  & - & - &  - & - & - &  -   \\
% (3j) & NCI or \multicolumn{2}{l}{DSI\textsubscript{d2q+daq} (PAWA + constrained + reg)}  & - & - &  - & - & - &  -   \\
\bottomrule
\end{tabular}
}
\caption{Results on small scale Natural Questions and TriviaQA datasets, reported in Recall@1 and Recall@5 respectively. First block presents baseline results in existing literature. Second block presents ablation results in a stacking fashion. Third block demonstrates the importance of document representation, in particular D2Q. Last row is the best method revised with in-domain D2Q.}
\label{table:v1:nci_small_scale}
\end{table*}

% in domain result
% (4a) & \multicolumn{2}{l}{NCI\textsubscript{daq + gen + d2q + idx + d2qid}} & 70.7 & 69.7 & 69.5 & 90.0 & 88.03 & 75.1 \\

\subsection{Implementation Details}
We use T5.1.1 as implemented by \texttt{t5x}~\cite{roberts2022t5x}.
We implement the different setups described in Section~\ref{sec.methods} in the form of \texttt{seqio} tasks. 
For the MS MARCO variants, we set the maximum input sequence length to 128 for all experiments, and  64 for the NQ100k and TriviaQA, following the NCI setup. We initialize our models with the pre-trained T5-base model. 
For the PAWA decoder, we randomly initialize the PAWA model parameters. 
Following \citep{Tay2022TransformerMA} for sequential IDs, beam search, with 40 beams, is used during inference.

We revise hyperparameter settings from \citep{Tay2022TransformerMA} to ones we have found to empirically perform better, especially for indexing larger corpora like MSMarcoFULL. 
We set the batch size in all our experiments to 512. 
We train our models with a learning rate of $10^{-3}$ and a dropout rate of $0.1$. 
We use 10k learning rate warm-up steps for all runs, except for Atomic IDs which use 100k steps.
We train our small-scale datasets,  NQ100k, TriviaQA, and MSMarco100k, for 1M steps.
For MSMarco1M and MSMarcoFULL, we train our model to convergence or, at most, 9M steps.
We use 8 TPUv4 chips for training models at the T5-Base scale.
T5-Large, T5-XL, and T5-Base with Atomic IDs over MSMarcoFULL use 64 TPUv4 chips. 
For T5-XXL, we use 128 chips.
Our most expensive runs took roughly 10-14 days to train to convergence on MSMarcoFULL.

\begin{table*}[t]
\centering
\resizebox{\textwidth}{!}{
\begin{tabular}{lllrrrrrrrrr}
\toprule
& & & \multicolumn{3}{c}{\textbf{MSMarco100k}} & \multicolumn{3}{c}{\textbf{MSMarco1M}} & \multicolumn{3}{c}{\textbf{MSMarcoFULL}}\\
\cmidrule(lr){4-6} \cmidrule(lr){7-9}  \cmidrule(lr){10-12} 
& \multicolumn{2}{l}{{\bf Model}} & At.\ & Nv.\ & Sm.\ & At.\ & Nv.\ & Sm.\ & At.\ & Nv.\ & Sm.\ \\
\toprule
\multicolumn{3}{l}{\emph{Baselines}}\\
\multicolumn{3}{l}{BM25} & - & 65.3 &  - & - & 41.3  & - & - & 18.4 & - \\
\multicolumn{3}{l}{BM25 (w/ doc2query--T5)} & - & 80.4 &  - & - & 56.6  & - & - & 27.2 & -\\
\multicolumn{3}{l}{GTR-Base} & - & 83.2 & - &  - & 60.7 & - & - & 34.8 & - \\
\midrule
\midrule
% \multicolumn{3}{l}{\emph{T5-base Variants}}\\ 
% (2a) & \multicolumn{2}{l}{FirstP only} & 0.0 & 6.7 & 5.6 &  0.0 & 1.2 & 1.8 & 0.0 & 10.8 & 1.2 \\
\multicolumn{3}{l}{\emph{Ours}}\\
(1a) & \multicolumn{2}{l}{Labeled Queries (No Indexing)} & 0.0 & 1.1 & 0.0 & 0.0 & 0.5 & 0.0 & 0.0 & 0.0 & 0.0 \\
(2a) & \multicolumn{2}{l}{FirstP/DaQ + Labeled Queries (DSI)} & 0.0 & 23.9 & 19.2 & 2.1 & 12.4  & 7.4 & 0.0 & 7.5 & 3.1 \\
(3b) & \multicolumn{2}{l}{FirstP/DaQ + D2Q + Labeled Queries} & 79.2 & 77.7 & 76.8 & 53.3 & 48.2 & 47.1 & 14.2 & \textbf{13.2} & 6.4 \\
% () & \multicolumn{2}{l}{FirstP + D2Q} & 79.6 & 79.1 & 78.3 & 52.0 & 53.2 & 52.4 & 22.6 & 14.7 & 6.8 \\

(4a) & \multicolumn{2}{l}{3b + PAWA (w/ 2D Semantic IDs)} & - & - & 77.1 & - & - & 50.2 & - & - & 9.0 \\
(5) & \multicolumn{2}{l}{4a + Consistency Loss (NCI)} & - & - & 77.1  & - & - & 50.2 & - & - &   9.1 \\
\midrule
(6b) & \multicolumn{2}{l}{D2Q only} & \textbf{80.3} & \textbf{78.7} & \textbf{78.5} & \textbf{55.8} & \textbf{55.4} & 54.0 & \textbf{24.2} & \textbf{13.3} & 11.8 \\
($4a^\prime$) & \multicolumn{2}{l}{6b + PAWA (w/ 2D Semantic IDs)} & - & - & 78.2 & - & - & \textbf{54.1} & - & - & \textbf{17.3} \\
($4b^\prime$) & \multicolumn{2}{l}{6b + Constrained Decoding} & - & - & \textbf{78.6} & - & - & 54.0 & - & - & 12.0 \\
($5^\prime$) & \multicolumn{2}{l}{6b + PAWA (w/ 2D Semantic IDs) + Constrained Decoding} & - & - & 78.3 & - & - & \textbf{54.2} & - & - & \textbf{17.4} \\
% \midrule
% \midrule
% \multicolumn{10}{c}{group align to table 2}\\
% (1a) gen only (partly , 2z)\\
% (2a) FirstP + gen (DSI) (\checkmark , 2b)  \\
% (3a) DaQ + gen + D2Q (\checkmark , 3a) \\
% (4a) 3b + pawa \\
% (4b) 3b + constrained\\
% (5) 3b + pawa + constrained (NCI) \\
% (2f) & \multicolumn{2}{l}{LOSS-CNCI\textsubscript{d2q}}  & 80.9 &  78.6 & 78.5   &   &  & \\
% (2g) & \multicolumn{2}{l}{PROMPT-CNCI\textsubscript{d2q}} & 79.6 & -  & -  &  & - & - &   & - & - \\
% (2h) & Hybrid(DSI-base, Baseline) \\
% \midrule
% \multicolumn{3}{l}{\emph{T5-XL Variants}}\\
% (3a) & \multicolumn{2}{l}{NCI\textsubscript{mix}} & $w_{2b}$ & $w_{2b}$  & $w_{2b}$  & $w_{2b}$  & $w_{2b}$ & $w_{2b}$  & $w_{2b}$  & $w_{2b}$ & $w_{2b}$ \\
% (3b) & \multicolumn{2}{l}{NCI\textsubscript{d2q}} &  &   &   &  &   &   &   &  &   \\
% (3c) & \multicolumn{2}{l}{LOSS-CNCI\textsubscript{d2q}}  & $w_{3b}$ & $w_{3b}$  & $w_{3b}$  & $w_{3b}$ &  $w_{3b}$ & $w_{3b}$  & $w_{3b}$  & $w_{3b}$ & $w_{3b}$\\
% (3d) & \multicolumn{2}{l}{PROMPT-CNCI\textsubscript{d2q}} & $w_{3b}$ & $w_{3b}$  & $w_{3b}$  & $w_{3b}$  & $w_{3b}$ & $w_{3b}$  & $w_{3b}$  & $w_{3b}$ & $w_{3b}$\\
% (3e) & Hybrid(DSI-XL, Baseline) \\
\bottomrule
\end{tabular}
}
\caption{Results on the development set of the scale variant MS MARCO V1 passage collections, reported in MRR@10. Best results per column and results within 0.1 of best are bolded. Note that FirstP here is equivalent to DaQ as MS MARCO input passages fit into the input window.}
\label{table:v1:passage_subset}
\end{table*}
\begin{table*}[t]
\centering
\resizebox{0.8\textwidth}{!}{
\begin{tabular}{llccc}
\toprule
\textbf{T5 Scale} & \textbf{Training} & \textbf{Params} & \textbf{Inference FLOPs} & \textbf{MRR@10} \\
\midrule
Base & D2Q Only + Atomic ID & 7.0B & $ 0.9 \times 10^{12}$ & 24.2 \\
Base & D2Q Only + Naive ID & 220M & $1.4 \times 10^{12}$ & 13.3 \\
Base & D2Q Only + PAWA (2D Sem.) & 761M & $ 6.8 \times 10^{12}$ & 17.3 \\
Large & D2Q Only + Naive ID & 783M & $3.5 \times 10^{12}$ & 21.4 \\
Large & D2Q Only + PAWA (2D Sem.) & 2.1B & $1.1 \times 10^{13}$ & 19.8 \\
XL & D2Q Only + Naive ID & 2.8B & $ 9.3 \times 10^{12}$ & \textbf{26.7} \\
XXL & D2Q Only + Naive ID & 11B & $ 4.3 \times 10^{13}$ & 24.3 \\
\bottomrule
\end{tabular}
}
\caption{Scaling up model size for sequential ID approaches in comparison to Atomic IDs for MSMarcoFULL.}
\label{table:v1:model_scale}
\end{table*}

\section{Experimental Results}\label{sec.results}

We report our results in three parts. First, we ablate all the methods from Section \ref{sec.methods} using T5-base on small-scale datasets: NQ100k and TriviaQA.
We observe which techniques work best on this small scale with widely studied datasets.
Then we transfer the same set of techniques and scale up to the entire MS MARCO passage ranking dataset to observe whether the same methods hold their ground at larger scales and discuss our findings.
Finally, to understand whether the effectiveness benefit from Atomic IDs can be attributed to additional model parameters on large corpora, we select the best approach and scale the model size up to 11B (T5-XXL equivalent) for sequential ID approaches.

\subsection{Ablations over Small Corpora}
We report our ablations over NQ100k and TriviaQA in Table \ref{table:v1:nci_small_scale}. The strongest combination of our techniques (row 7) sets a new state-of-the-art result on NCI's variant of NQ, without using any sophisticated modeling techniques such as architecture changes or learned docids.

The choice of document representation by far dominates the overall performance of the retriever. 
Using just the training queries provided by the dataset performs the worst due to the low coverage of the documents.
FirstP is a major improvement over this and DaQ is better than FirstP.
However, the usage of D2Q is essential to strong generative retrieval performance, resulting in a 7pt+ gain.
This by far trumps all other proposed techniques.

As for other design choices, we see that at this small scale naive and Semantic IDs perform about on par (varying between task configurations), with Atomic IDs consistently the best. We note though that on NQ100k, Atomic IDs add 80M parameters to a T5-Base model that would otherwise be 220M parameters (a $36\%$ increase). Given the comparable performance in the best configuration (row 7), these extra parameters may or may not be worth it, but we refer to Section \ref{sec.discussion-scaling} for more discussion. Modeling techniques from~\cite{Wang2022ANC}, i.e. 2D Semantic IDs, PAWA, constrained decoding, and consistency loss, do not reliably improve the model over the use of synthetic queries alone.

At this corpus scale, our best result uses a mixture of FirstP, DaQ, labeled queries, and synthetic queries for training. However, importantly, the \textit{quality} of the synthetic queries are quite important, with synthetic queries from a generator specifically trained for the question answering domain significantly outperforming the query generator trained over MS MARCO which was used by previous works.  

\subsection{Scaling Corpus Size}

We now consider the scaled version of the MS MARCO passage ranking task, scaling from 100k to 1M and 8.8M passages. Results are reported in Table \ref{table:v1:passage_subset}.
Perhaps the most striking observation about the transition to MS MARCO is the absolute requirement of synthetic queries for strong retrieval performance. Synthetic queries result in a 2-3x improvement over the original DSI formulation alone. In fact, using only synthetic queries to docid as the indexing task is the most effective and straightforward training strategy on MS MARCO. This is a notable difference in the transition from NQ and TriviaQA to MS MARCO, where FirstP and DaQ did provide substantial value. This may be due to NQ and TriviaQA being based on Wikipedia articles: the beginning of Wikipedia documents are informative entity descriptions, and many sentences refer to the entity--which is likely the answer to a requested query. 

As corpus size grows, DSI performance rapidly drops off, with the best result (D2Q only with Atomic IDs) rapidly falling off from 80.3 to 55.8 and finally 24.2 as we scale to the full 8.8M passages. Vanilla Semantic IDs also drop off as we scale to the full corpus, under-performing naive identifiers. We conjecture that this may be due to the potentially increased length of semantic identifiers being more difficult to decode than naive identifiers coupled with a noisy partitioning of the semantic space (especially when using an off-the-shelf embedding model such as SentenceT5-Base.) However, we do observe that Semantic IDs decoded via PAWA perform better. We provide some insight into why this might be in the next section where we examine model size. Constrained decoding only provides marginal value and generally is not worth the added complexity.

\subsection{Scaling Model Size}
How much of Atomic ID's strong performance can be attributed to its additional model parameters? On MSMarcoFULL, decoding Atomic ID document tokens adds an additional ~7B parameters to the otherwise 220M parameter T5-Base model. We take the best configuration on MSMarcoFULL from Table \ref{table:v1:passage_subset} and scale model parameters of Naive ID and Semantic ID (PAWA) to similar sizes for comparison. We report results in Table \ref{table:v1:model_scale}.

Overall, we observe a general trend that as parameter count increases, retrieval performance improves. Indeed, both Atomic IDs and PAWA Semantic IDs had the strongest performance in Table \ref{table:v1:passage_subset}, which we now attribute to their increased size. Notice that the difference here only comes out when scaling to MSMarcoFULL, where these parameter differences magnify significantly over smaller corpus scales.

However, not all methods are equal. PAWA and 2D Semantic IDs \citep{Wang2022ANC} significantly increase decoding parameters with its extra decoding stack, yet yield no gain over naively scaling the Transformer with Naive IDs, underperforming by 4pts at around ~700M parameters. This pattern continues to hold scaling PAWA to 2.1B parameters, thus, in order to save resources, we do not scale PAWA any further.

Scaling Transformers naively according to default T5 scales while using Naive IDs had the strongest performance on MSMarcoFULL at 26.7 MRR@10. Using only ~2.8B parameters, this approach outperforms T5-Base with Atomic IDs  which uses 7B parameters while achieving only 24.2 MRR@10. However, while parameter count has practical implications regarding the resources required for training and inference (especially TPU/GPU memory), there are other trade-offs to consider, which we discuss in the next section.

While Naive IDs perform well at T5-XL size, surprisingly we find that scaling further to XXL (11B) does not improve performance; in fact, it is detrimental to retrieval performance (24.3 MRR@10 vs. XL's 26.7) under the same experimental settings and hyper-parameter settings, even though model training converges faster. This is counter-intuitive to most generative tasks and to the typical intuition of generative retrieval relying on model capacity to index an entire corpus of documents.

\section{Discussion}
\label{sec.discussion}
% Finally, we try to answer R3 through empirical studies.
% Unfortunately, there exist no short answer for this. 
The results of this work raises multiple questions regarding the current state of generative retrieval at scale which we aim to provide more insight here.

\subsection{Why are synthetic queries effective?}\label{sec.discussion-queries} Although the use of synthetic queries as a document representation technique has been shown to be effective in previous works \citep{zhuang2022bridging, Wang2022ANC, chen2023understanding}, our experiments highlight its central importance to generative retrieval on a larger, more challenging corpus. We suggest that the effectiveness of synthetic queries mainly come from augmenting the input distribution during training to be closer to that observed at inference/evaluation time. Mainly, this comes in two forms: mitigating the coverage gap of ground-truth labeled queries and the document corpus, and closing the gap between the training query distribution and inference/evaluation. In addition, we find that the diversity of generated synthetic queries also can have a significant effect on retrieval performance.
\\\\
% D2Q-40 - Best Eval - /cns/yo-d/home/unicorn-ai/dsi/models/ronak/paper/nci/dfx_msmarco_passage_d2q_ND40_K10_TEM100_0words_tiny_t5base_atomic_van__02-12-22-55/inference_eval/dfx_msmarco_passage_gen_eval_only_tiny_spm_en_c4_dx100k-1160000.jsonl
% KendalTau - SignificanceResult(statistic=0.2459251531187779, pvalue=6.02383785869447e-145)
% PearsonRResult(statistic=0.31702373322351995, pvalue=9.58667684477144e-163)
% SpearmanR - SignificanceResult(statistic=0.3042749763961819, pvalue=1.9233677238761753e-149)
% D2Q-100 - /cns/yo-d/home/unicorn-ai/dsi/models/ronak/paper/nci/dfx_msmarco_passage_d2q_ND100_K10_TEM100_0words_tiny_t5base_atomic_van__02-12-22-53/inference_eval/dfx_msmarco_passage_gen_eval_only_tiny_spm_en_c4_dx100k-1250000.jsonl
% KendalTau - SignificanceResult(statistic=0.2630753879627397, pvalue=4.889429829246177e-160)
% PearsonRResult(statistic=0.342775845290883, pvalue=1.1254445649572709e-191)
% SpearmanR -SignificanceResult(statistic=0.3194502414990135, pvalue=2.368547887324849e-165)
\begin{figure}[t]
    \centering
\begin{tikzpicture}
\begin{axis}[
    width=0.49\textwidth,
    height=0.4\textwidth,
    scaled x ticks=false,
    scaled y ticks=false,
    xlabel={Jaccard Similarity (\%)},
    ylabel={MRR@10},
    symbolic x coords={10, 20, 30, 40, 50, 60, 70, 80, 90, 100},
    xtick=data,
    ytick={70,80,90,100},
    legend pos=north west,
    enlargelimits=true,
    grid style=dashed,
    legend pos=south east,
    legend style={nodes={scale=0.6, transform shape}},
]

\addplot[
    color=gray,
    mark=square,
    ]
    coordinates {
    (10,73.49)(20,78.27)(30,83.85)(40,86.02)(50,86.29)(60,89.55)(70,88.75)(80,93.33)(90,97.08)(100,95.36)
    };
    \addlegendentry{D2Q-100}
    
\addplot[
    color=gray,
    mark=triangle,
    ]
    coordinates {
    (10,70.13)(20,75.63)(30,80.4)(40,83.68)(50,84.71)(60,86.81)(70,89.39)(80,88.90)(90,90.10)(100,96.54)
    };
    \addlegendentry{D2Q-40}

\end{axis}
\end{tikzpicture}
\caption{Jaccard similarity between synthetic queries and validation set queries vs.\ MRR@10 on the MSMarco100K subset.}
\label{fig:jaccard_mrr}
\end{figure}
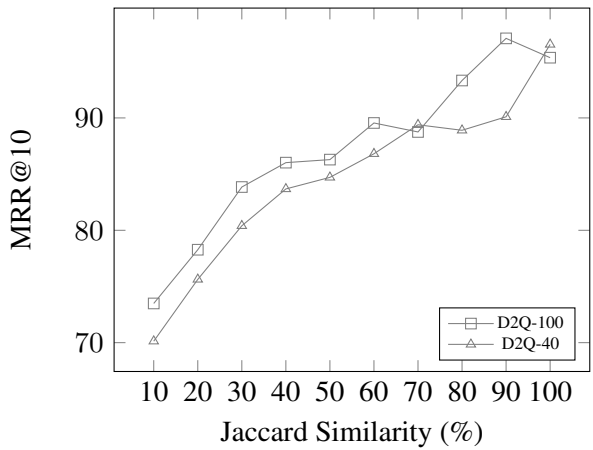
% https://tensorboard.corp.google.com/compare/qfiltered10:6241548422347635571,rfiltered10:1762866038444156087,rfiltered30:4707610559598456763,qfiltered30:2828400863249744305,rfiltered20:6498115402992944315,qfiltered20:5445932475994253836,rfiltered40:1131740095436489871,qfiltered40:4872914575756092330/?darkMode=true&forceSVG=true#timeseries

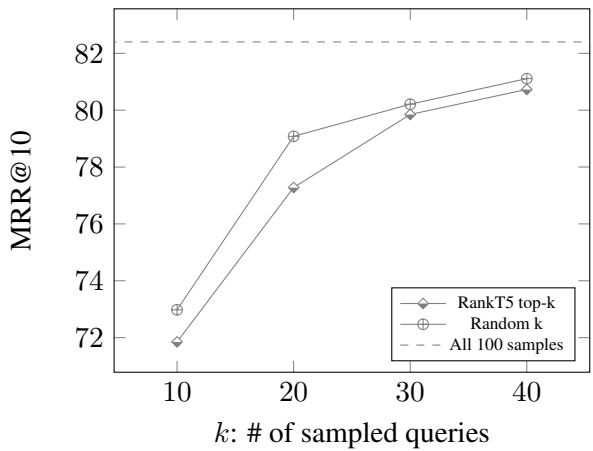
\begin{figure}[t]
    \centering
\begin{tikzpicture}
\begin{axis}[
    width=0.49\textwidth,
    height=0.4\textwidth,
    scaled x ticks=false,
    scaled y ticks=false,
    xlabel={$k$: \# of sampled queries},
    ylabel={MRR@10},
    %symbolic x coords={10, 20, 30, 40},
    xmin=8,
    xmax=42,
    xtick=data,
    ytick={70, 72, 74, 76, 78, 80, 82, 84},
    ymax=82.5,
    legend pos=north west,
    enlargelimits=true,
    grid style=dashed,
    legend pos=south east,
    legend style={nodes={scale=0.6, transform shape}},
]

\addplot[
    color=gray,
    mark=halfsquare*,
    ]
    coordinates {
    (10,71.85)(20,77.28)(30,79.85)(40,80.73)
    };
    \addlegendentry{RankT5 top-k}
    
\addplot[
    color=gray,
    mark=oplus,
    ]
    coordinates {
    (10,72.98)(20,79.08)(30,80.21)(40,81.11)
    };
    \addlegendentry{Random k}
    
\addplot[
    color=gray,
    dashed
    ]
    coordinates {
    (0, 82.4)(50, 82.4)
    };
    \addlegendentry{All 100 samples}

\end{axis}
\end{tikzpicture}
\caption{MSMarco100K MRR@10 as we vary the number of synthetic queries per passage. Given 100 pre-generated queries per passage, we compare random-k sampling, top-k selection via RankT5-XL, and using all 100 synthetic queries.  }
\label{fig:rankt5}
\end{figure}
\paragraph{Document coverage gap.} In Table \ref{tab.dataset}, for each dataset we report the coverage of their document corpus by the corresponding labeled query training set. When comparing MSMarco100k, 1M, and FULL the query coverage drops from 92.9\% to 51.6\% and 5.8\% respectively. Consider rows (2a) and (3b) in Table \ref{table:v1:passage_subset} which only differ by the addition of synthetic queries. Here we observe that MSMarco100k improved by 3.3x while MSMarco1M improved by 3.9x, even though 1M is a larger corpus and may be affected by model capacity as we see with MSMarcoFULL. Similarly, for NQ100k and TriviaQA, which have 98.4\% and 57.7\% coverage respectively, we observe that swapping Labeled Queries (No Indexing) (row 1a) for D2Q only (row 6b) hurts performance for NQ100k while improving performance for TriviaQA (Table \ref{table:v1:nci_small_scale}.) Since this D2Q model is trained on MS MARCO, for NQ100k replacing its own labeled queries with synthetic queries only amounted to a 1.6\% coverage gain, which is not worth the domain shift. However, for TriviaQA this amounted to a 42.3\% coverage gain, which is more worth the domain shift.
\\\\
\paragraph{Query distribution gap.} Synthetic query generation effectively closes the query distribution gap between training and evaluation. Table \ref{table:v1:nci_small_scale} row 7, first shows the importance of the query distribution by using an in-domain query generation model to improve retrieval performance. To further understand the relationship between retrieval performance and query distribution gap, we plot the relationship between synthetic query similarity vs. validation query similarity and retrieval performance (MRR@10). For each evaluation query in the MS MARCO validation set, we measure the maximum similarity among all synthetic queries generated for the corresponding passage. Jaccard similarity is used for simplicity. For each evaluation query we then evaluate MRR@10 using the Atomic ID variant of row 6b in Table \ref{table:v1:passage_subset}. Figure \ref{fig:jaccard_mrr} reports the average MRR@10 within each 10pt Jaccard similarity bucket. We plot two variants using 40 and 100 sampled queries per passage for comparison.

In general, higher Jaccard similarity correlates with higher MRR@10 performance. That is, the more similar our training queries are to the evaluation the stronger the retrieval performance. Comparing the two settings, we see that higher exposure to more synthetic queries typically promotes higher effectiveness across similarity buckets. Even though the query distribution is important, it is worth noting that even on the lowest end of similarity this setting still has strong retrieval performance. While synthetic query distribution is an important aspect of retrieval performance, it is not singular in determining the end effectiveness and the generative retrieval model goes far beyond simply detecting lexically similar queries to those seen during training.
\\\\
\paragraph{Diversity.} We provide further analysis regarding the importance of synthetic query diversity. Here we assume the same MSMarco100k setting using the Atomic ID variant of row 6b in Table \ref{table:v1:passage_subset}. We vary the number of sampled synthetic queries per passage used for training and observe MRR@10. We consider using 10, 20, 30, 40 and 100 sampled queries per passage, which we construct by first sampling the full 100 then taking random subsets of the varying sizes. We use a sampling temperature of 1.0 and consider the top 10 tokens at each sampling step. Recent studies show advances in utilizing cross encoders to refine the generated query set of incoherent, unspecific queries to improve the use of D2Q \citep{d2q-less-more}. Accordingly, we also experiment with ranking the 100 sampled queries and taking top-10,20,30,40 instead of randomly sampling. We do so using a state-of-the-art cross-attention re-ranker, RankT5-XL~\cite{zhuang22rankt5}, to score (generated query, passage) pairs and then take the top-k. 

We report results in Figure \ref{fig:rankt5}. We find that, consistently, sampling more synthetic queries improve performance in this setting. Surprisingly, applying RankT5-based selection over the samples hurt performance. This suggests an overall preference for more samples, and more diverse samples to improve effectiveness. Using all 100 samples performed the best, increasing MRR@10 from 80.3 (Table \ref{table:v1:passage_subset}, which used 40 samples) to 82.4, closing the gap with GTR-Base (83.2 MRR@10) on MSMarco100k. Exactly why query diversity is so important still up for interpretation, but there could be a couple possiblities: more diverse samples gives higher probability of at least some of the samples being close to the target distribution and more samples could provide a type of regularization to the model.

\subsection{Which model scaling approach is best?}\label{sec.discussion-scaling} Much of this paper has considered parameter cost as a proxy for memorization capacity, which has been conjectured in the past to be important for retrieval \citep{Tay2022TransformerMA}. However, model comparisons should not stop at parameter counts as this may not correlate with other cost indicators (training speed, FLOPs, etc.) that are important to practical applications \citep{dehghani2022the}. While ultimately the best method to scale generative retrieval models will be the one that unlocks the potential of the paradigm to be competitive on large scale retrieval tasks, we can provide some first glimpses into what trade-offs are at stake as we consider larger models for larger corpora.

As a case study, we consider T5-Base with Atomic IDs compared as T5-XL with Naive IDs from Table \ref{table:v1:model_scale}. Both are trained only with synthetic queries, and represent the only two viable approaches from our experiments. PAWA severely underperforms with regards to quality as we scale model size, not to mention the FLOP expense of having an extra decoding stack during inference. We provide discussion on parameter cost, training speed, and inference FLOPs here.

\textit{Parameters.} As corpus size scales, generative retrieval models face a fundamental prerequisite in model size to achieve decent performance, as seen in Table \ref{table:v1:passage_subset}. Between three different ways of adding parameters (naive scaling, Atomic IDs, PAWA decoder), we see quality improvements over the smaller models. As discussed, on a fixed parameter budget basis Naive IDs perform the best on MSMarcoFULL, and best in quality overall.

\textit{Training Speed.} Applications that require frequent retraining value fast total training walltime. We train T5-Base Atomic IDs and T5-XL Naive IDs on the same hardware (64 TPUv4) and hyperparameter settings. To achieve the optimal performance reported in Table \ref{table:v1:model_scale}, T5-XL Naive IDs required ~14 days while T5-Base Atomic ID required only ~7 days. However, at ~7 days T5-XL Naive IDs was quality matched with T5-Base Atomic IDs (~24.5 MRR@10), making both approaches roughly equal in terms of training wall-time when accounting for quality. 

\textit{Inference FLOPs.} Inference FLOPs can be a proxy for serving performance, although imperfect. Here we see that while sequential identifiers can achieve more with fewer parameters, atomic identifiers are incredibly FLOP efficient during inference. T5-Base with Atomic IDs for MSMarcoFULL requires only 9.7\% the inference FLOPs of T5-XL with Naive IDs for 90\% of the retrieval performance (Table \ref{table:v1:model_scale}). How is this possible? Atomic IDs incur additional compute cost to compute an output projection and softmax over the enormous vocab of ~8.8M docids. However, it only has to compute this once to get a complete ranking of the \textit{entire} corpus -- a potentially very special property of the approach. On the other hand, sequential identifiers require $d$ decoding steps to decode a single docid, and $k$ beams to find a ranking of $k$ docids. $k=40$ for our experiments. Thus even though Atomic IDs require an expensive output projection, sequential ids require $O(d \cdot k)$ more decoding steps. To scale Naive IDs to be competitive with Atomic IDs, also makes individual decoding steps significantly more expensive.

In the end, we cannot yet say which approach is the best as the paradigm has yet to achieve competitive results on MS Marco passage ranking. On small corpora (100k), Atomic IDs are the highest quality, efficient option without incurring too many extra parameters. From our experiments though we can see that training models to maximize memorization amplifies compute trade-offs, and the field must provide more nuanced discussions of cost trade-offs as it considers more realistic applications of generative retrieval.

\section{Limitations}
As with all empirical studies, ours has its own set of limitations which we urge the reader to consider.
Multiple works have come after the experiments in this work, e.g., \cite{chen2023understanding}, and thus we do not present an exhaustive set of generative retrieval techniques here. For example, the wide space of identifiers based on natural language or learned codes.
In addition, due to resource constraints our model scaling experiments are not exhaustive, and not all ablation scenarios in Table \ref{table:v1:passage_subset} are scaled to larger model sizes. It could be possible that certain setups improve more at larger parameterizations, although unlikely; as with scaling past 11B.
In addition, due to the extreme parameter requirements we do not saturate the scaling of Atomic IDs.
%Finally, while this work provides insight into the effectiveness of generative retrieval on large corpora, it does not answer the question of: what is the largest possible corpus where generative retrieval could be the state-of-the-art on? I.e., scaling model size for smaller corpora. This is a different question of practical importance which we leave for future work.
Finally, since this work focused on the effectiveness of generative retrieval on large corpora, scaling model size for smaller corpora was outside our scope. Investigating the maximum corpus size for which generative retrieval could provide state-of-the-art performance is a question of practical importance which we leave for future work.

\section{Future Directions}\label{sec.future}
While open problems in generative retrieval have not changed (e.g. how to achieve state-of-the-art results on large corpora, how to update such as model with new documents \citep{mehta2022dsi}, etc), we believe that our work also raises new  open questions for the field. (1) How do we properly leverage large language models and the power of scaling model parameters to benefit generative retrieval on large corpora? While \citet{Tay2022TransformerMA} showed this possibility over NQ, the same is not yet observed on MS MARCO even though intuitively expanded model capacity \textit{should} benefit increased corpus scale. (2) How can we design model scaling recipes and derive scaling laws that maximize retrieval performance? In this work we only consider default T5 parameterizations, which may or may not be optimal for memorization heavy tasks. (3) How can we design architectures that can interpolate between the compute trade-offs of Atomic IDs and sequential IDs? We look forward to understand more about these problems in future works.

\section{Conclusion}\label{sec.conclusion}

We provide the first empirical study of generative retrieval methods over the full MS MARCO passage ranking task of 8.8M passages. Of the various methods from the literature which we consider in this work \citep{Tay2022TransformerMA, zhuang2022bridging, Wang2022ANC}, we find that the use of synthetic queries as a document representation strategy is the only approach that remained effective, and necessary, as we scaled up the corpus size using MS MARCO passages. We also highlight the importance of accounting for the compute cost of techniques; keeping the parameter count fixed, we find that naive methods outperform more sophisticated ones on the full MS MARCO dataset. Our strongest result on MS MARCO passage ranking uses only synthetic queries to Naive IDs as its training task, with the model scaled to T5-XL (3B). This model achieves 26.7 MRR@10. Surprisingly, increasing parameters for the same setting up to XXL (11B) performs worse. All of these findings suggest a need for continued research into generative retrieval, closer attention to method comparisons, and the potential need for fundamental improvements to the paradigm before we can leverage the power of larger language models.

\section{Acknowledgements}
The authors would like to thank Yi Tay, Tal Schuster, and Sanket Vaibhav Mehta for their valuable feedback and discussions.

% \begin{acks}
% \end{acks}

% \clearpage
%\balancecolumns

\bibliographystyle{ACM-Reference-Format}
\bibliography{ranking}


\begin{thebibliography}{46}

%%% the Web address.

\ifx \showCODEN    \undefined \def \showCODEN     #1{\unskip}     \fi
\ifx \showDOI      \undefined \def \showDOI       #1{#1}\fi
\ifx \showISBNx    \undefined \def \showISBNx     #1{\unskip}     \fi
\ifx \showISBNxiii \undefined \def \showISBNxiii  #1{\unskip}     \fi
\ifx \showISSN     \undefined \def \showISSN      #1{\unskip}     \fi
\ifx \showLCCN     \undefined \def \showLCCN      #1{\unskip}     \fi
\ifx \shownote     \undefined \def \shownote      #1{#1}          \fi
\ifx \showarticletitle \undefined \def \showarticletitle #1{#1}   \fi
\ifx \showURL      \undefined \def \showURL       {\relax}        \fi
% invisible to TeX
\providecommand\bibfield[2]{#2}
\providecommand\bibinfo[2]{#2}
\providecommand\natexlab[1]{#1}
\providecommand\showeprint[2][]{arXiv:#2}

\bibitem[\protect\citeauthoryear{Asai, Kasai, Clark, Lee, Choi, and
  Hajishirzi}{Asai et~al\mbox{.}}{2021}]%
        {asai-etal-2021-xor}
\bibfield{author}{\bibinfo{person}{Akari Asai}, \bibinfo{person}{Jungo Kasai},
  \bibinfo{person}{Jonathan Clark}, \bibinfo{person}{Kenton Lee},
  \bibinfo{person}{Eunsol Choi}, {and} \bibinfo{person}{Hannaneh Hajishirzi}.}
  \bibinfo{year}{2021}\natexlab{}.
\newblock \showarticletitle{{XOR} {QA}: Cross-lingual Open-Retrieval Question
  Answering}. In \bibinfo{booktitle}{\emph{Proceedings of the 2021 Conference
  of the North American Chapter of the Association for Computational
  Linguistics: Human Language Technologies}}. \bibinfo{publisher}{Association
  for Computational Linguistics}, \bibinfo{address}{Online},
  \bibinfo{pages}{547--564}.
\newblock
\urldef\tempurl%
\url{https://doi.org/10.18653/v1/2021.naacl-main.46}
\showDOI{\tempurl}


\bibitem[\protect\citeauthoryear{Bevilacqua, Ottaviano, Lewis, Yih, Riedel, and
  Petroni}{Bevilacqua et~al\mbox{.}}{2022}]%
        {autoregressive-search-engine}
\bibfield{author}{\bibinfo{person}{Michele Bevilacqua},
  \bibinfo{person}{Giuseppe Ottaviano}, \bibinfo{person}{Patrick Lewis},
  \bibinfo{person}{Wen-tau Yih}, \bibinfo{person}{Sebastian Riedel}, {and}
  \bibinfo{person}{Fabio Petroni}.} \bibinfo{year}{2022}\natexlab{}.
\newblock \bibinfo{title}{Autoregressive Search Engines: Generating Substrings
  as Document Identifiers}.
\newblock
\newblock
\urldef\tempurl%
\url{https://doi.org/10.48550/ARXIV.2204.10628}
\showDOI{\tempurl}


\bibitem[\protect\citeauthoryear{Chen, Liu, He, Sun, and Sun}{Chen
  et~al\mbox{.}}{2023}]%
        {chen2023understanding}
\bibfield{author}{\bibinfo{person}{Xiaoyang Chen}, \bibinfo{person}{Yanjiang
  Liu}, \bibinfo{person}{Ben He}, \bibinfo{person}{Le Sun}, {and}
  \bibinfo{person}{Yingfei Sun}.} \bibinfo{year}{2023}\natexlab{}.
\newblock \showarticletitle{Understanding Differential Search Index for Text
  Retrieval}.
\newblock \bibinfo{journal}{\emph{arXiv preprint arXiv:2305.02073}}
  (\bibinfo{year}{2023}).
\newblock


\bibitem[\protect\citeauthoryear{Chen, Luo, He, Sun, and Sun}{Chen
  et~al\mbox{.}}{2022}]%
        {chen2022towards}
\bibfield{author}{\bibinfo{person}{Xuanang Chen}, \bibinfo{person}{Jian Luo},
  \bibinfo{person}{Ben He}, \bibinfo{person}{Le Sun}, {and}
  \bibinfo{person}{Yingfei Sun}.} \bibinfo{year}{2022}\natexlab{}.
\newblock \showarticletitle{Towards robust dense retrieval via local ranking
  alignment}. In \bibinfo{booktitle}{\emph{Proceedings of the Thirty-First
  International Joint Conference on Artificial Intelligence, IJCAI}}.
  \bibinfo{pages}{1980--1986}.
\newblock


\bibitem[\protect\citeauthoryear{Craswell, Mitra, Yilmaz, Campos, and
  Lin}{Craswell et~al\mbox{.}}{2022}]%
        {craswell2022overview}
\bibfield{author}{\bibinfo{person}{Nick Craswell}, \bibinfo{person}{Bhaskar
  Mitra}, \bibinfo{person}{Emine Yilmaz}, \bibinfo{person}{Daniel Campos},
  {and} \bibinfo{person}{Jimmy Lin}.} \bibinfo{year}{2022}\natexlab{}.
\newblock \showarticletitle{Overview of the TREC 2022 deep learning track}. In
  \bibinfo{booktitle}{\emph{Text REtrieval Conference (TREC)}}.
  \bibinfo{publisher}{TREC}.
\newblock
\urldef\tempurl%
\url{https://www.microsoft.com/en-us/research/publication/overview-of-the-trec-2021-deep-learning-track/}
\showURL{%
\tempurl}


\bibitem[\protect\citeauthoryear{De~Cao, Izacard, Riedel, and Petroni}{De~Cao
  et~al\mbox{.}}{2020}]%
        {cao-aer}
\bibfield{author}{\bibinfo{person}{Nicola De~Cao}, \bibinfo{person}{Gautier
  Izacard}, \bibinfo{person}{Sebastian Riedel}, {and} \bibinfo{person}{Fabio
  Petroni}.} \bibinfo{year}{2020}\natexlab{}.
\newblock \bibinfo{title}{Autoregressive Entity Retrieval}.
\newblock
\newblock
\urldef\tempurl%
\url{https://doi.org/10.48550/ARXIV.2010.00904}
\showDOI{\tempurl}


\bibitem[\protect\citeauthoryear{Dehghani, Tay, Arnab, Beyer, and
  Vaswani}{Dehghani et~al\mbox{.}}{2022}]%
        {dehghani2022the}
\bibfield{author}{\bibinfo{person}{Mostafa Dehghani}, \bibinfo{person}{Yi Tay},
  \bibinfo{person}{Anurag Arnab}, \bibinfo{person}{Lucas Beyer}, {and}
  \bibinfo{person}{Ashish Vaswani}.} \bibinfo{year}{2022}\natexlab{}.
\newblock \showarticletitle{The Efficiency Misnomer}. In
  \bibinfo{booktitle}{\emph{International Conference on Learning
  Representations}}.
\newblock
\urldef\tempurl%
\url{https://openreview.net/forum?id=iulEMLYh1uR}
\showURL{%
\tempurl}


\bibitem[\protect\citeauthoryear{Devlin, Chang, Lee, and Toutanova}{Devlin
  et~al\mbox{.}}{2019}]%
        {devlin2019bert}
\bibfield{author}{\bibinfo{person}{Jacob Devlin}, \bibinfo{person}{Ming-Wei
  Chang}, \bibinfo{person}{Kenton Lee}, {and} \bibinfo{person}{Kristina
  Toutanova}.} \bibinfo{year}{2019}\natexlab{}.
\newblock \showarticletitle{{BERT}: Pre-training of Deep Bidirectional
  Transformers for Language Understanding}. In
  \bibinfo{booktitle}{\emph{NAACL-HLT (1)}}.
\newblock


\bibitem[\protect\citeauthoryear{Gao, Yao, and Chen}{Gao et~al\mbox{.}}{2021}]%
        {gao-etal-2021-simcse}
\bibfield{author}{\bibinfo{person}{Tianyu Gao}, \bibinfo{person}{Xingcheng
  Yao}, {and} \bibinfo{person}{Danqi Chen}.} \bibinfo{year}{2021}\natexlab{}.
\newblock \showarticletitle{{S}im{CSE}: Simple Contrastive Learning of Sentence
  Embeddings}. In \bibinfo{booktitle}{\emph{Proceedings of the 2021 Conference
  on Empirical Methods in Natural Language Processing}}.
  \bibinfo{publisher}{Association for Computational Linguistics},
  \bibinfo{address}{Online and Punta Cana, Dominican Republic},
  \bibinfo{pages}{6894--6910}.
\newblock
\urldef\tempurl%
\url{https://doi.org/10.18653/v1/2021.emnlp-main.552}
\showDOI{\tempurl}


\bibitem[\protect\citeauthoryear{Gillick, Presta, and Tomar}{Gillick
  et~al\mbox{.}}{2018}]%
        {gillick2018end}
\bibfield{author}{\bibinfo{person}{Daniel Gillick}, \bibinfo{person}{Alessandro
  Presta}, {and} \bibinfo{person}{Gaurav~Singh Tomar}.}
  \bibinfo{year}{2018}\natexlab{}.
\newblock \showarticletitle{End-to-end retrieval in continuous space}.
\newblock \bibinfo{journal}{\emph{arXiv preprint arXiv:1811.08008}}
  (\bibinfo{year}{2018}).
\newblock


\bibitem[\protect\citeauthoryear{Gospodinov, MacAvaney, and
  Macdonald}{Gospodinov et~al\mbox{.}}{2023}]%
        {d2q-less-more}
\bibfield{author}{\bibinfo{person}{Mitko Gospodinov}, \bibinfo{person}{Sean
  MacAvaney}, {and} \bibinfo{person}{Craig Macdonald}.}
  \bibinfo{year}{2023}\natexlab{}.
\newblock \showarticletitle{Doc2Query--: When Less is More}.
\newblock  (\bibinfo{year}{2023}).
\newblock
\urldef\tempurl%
\url{https://doi.org/10.48550/ARXIV.2301.03266}
\showDOI{\tempurl}


\bibitem[\protect\citeauthoryear{Hadsell, Chopra, and LeCun}{Hadsell
  et~al\mbox{.}}{2006}]%
        {Hadsell2006DimensionalityRB}
\bibfield{author}{\bibinfo{person}{Raia Hadsell}, \bibinfo{person}{Sumit
  Chopra}, {and} \bibinfo{person}{Yann LeCun}.}
  \bibinfo{year}{2006}\natexlab{}.
\newblock \showarticletitle{Dimensionality Reduction by Learning an Invariant
  Mapping}.
\newblock \bibinfo{journal}{\emph{2006 IEEE Computer Society Conference on
  Computer Vision and Pattern Recognition (CVPR'06)}}  \bibinfo{volume}{2}
  (\bibinfo{year}{2006}), \bibinfo{pages}{1735--1742}.
\newblock


\bibitem[\protect\citeauthoryear{Hui, Zhuang, Chen, Qin, Lu, Bahri, Ma, Gupta,
  Nogueira~dos Santos, Tay, and Metzler}{Hui et~al\mbox{.}}{2022}]%
        {hui-etal-2022-ed2lm}
\bibfield{author}{\bibinfo{person}{Kai Hui}, \bibinfo{person}{Honglei Zhuang},
  \bibinfo{person}{Tao Chen}, \bibinfo{person}{Zhen Qin}, \bibinfo{person}{Jing
  Lu}, \bibinfo{person}{Dara Bahri}, \bibinfo{person}{Ji Ma},
  \bibinfo{person}{Jai Gupta}, \bibinfo{person}{Cicero Nogueira~dos Santos},
  \bibinfo{person}{Yi Tay}, {and} \bibinfo{person}{Donald Metzler}.}
  \bibinfo{year}{2022}\natexlab{}.
\newblock \showarticletitle{{ED}2{LM}: Encoder-Decoder to Language Model for
  Faster Document Re-ranking Inference}. In \bibinfo{booktitle}{\emph{Findings
  of the Association for Computational Linguistics: ACL 2022}}.
  \bibinfo{publisher}{Association for Computational Linguistics},
  \bibinfo{address}{Dublin, Ireland}, \bibinfo{pages}{3747--3758}.
\newblock
\urldef\tempurl%
\url{https://doi.org/10.18653/v1/2022.findings-acl.295}
\showDOI{\tempurl}


\bibitem[\protect\citeauthoryear{Johnson, Douze, and Jégou}{Johnson
  et~al\mbox{.}}{2021}]%
        {JonhsonKNN}
\bibfield{author}{\bibinfo{person}{Jeff Johnson}, \bibinfo{person}{Matthijs
  Douze}, {and} \bibinfo{person}{Hervé Jégou}.}
  \bibinfo{year}{2021}\natexlab{}.
\newblock \showarticletitle{Billion-Scale Similarity Search with GPUs}.
\newblock \bibinfo{journal}{\emph{IEEE Transactions on Big Data}}
  \bibinfo{volume}{7}, \bibinfo{number}{3} (\bibinfo{year}{2021}),
  \bibinfo{pages}{535--547}.
\newblock
\urldef\tempurl%
\url{https://doi.org/10.1109/TBDATA.2019.2921572}
\showDOI{\tempurl}


\bibitem[\protect\citeauthoryear{Joshi, Choi, Weld, and Zettlemoyer}{Joshi
  et~al\mbox{.}}{2017}]%
        {joshi2017triviaqa}
\bibfield{author}{\bibinfo{person}{Mandar Joshi}, \bibinfo{person}{Eunsol
  Choi}, \bibinfo{person}{Daniel~S Weld}, {and} \bibinfo{person}{Luke
  Zettlemoyer}.} \bibinfo{year}{2017}\natexlab{}.
\newblock \showarticletitle{Triviaqa: A large scale distantly supervised
  challenge dataset for reading comprehension}.
\newblock \bibinfo{journal}{\emph{arXiv preprint arXiv:1705.03551}}
  (\bibinfo{year}{2017}).
\newblock


\bibitem[\protect\citeauthoryear{Karpukhin, Oguz, Min, Lewis, Wu, Edunov, Chen,
  and Yih}{Karpukhin et~al\mbox{.}}{2020}]%
        {karpukhin2020dense}
\bibfield{author}{\bibinfo{person}{Vladimir Karpukhin}, \bibinfo{person}{Barlas
  Oguz}, \bibinfo{person}{Sewon Min}, \bibinfo{person}{Patrick Lewis},
  \bibinfo{person}{Ledell Wu}, \bibinfo{person}{Sergey Edunov},
  \bibinfo{person}{Danqi Chen}, {and} \bibinfo{person}{Wen-tau Yih}.}
  \bibinfo{year}{2020}\natexlab{}.
\newblock \showarticletitle{Dense Passage Retrieval for Open-Domain Question
  Answering}. In \bibinfo{booktitle}{\emph{Proceedings of the 2020 Conference
  on Empirical Methods in Natural Language Processing (EMNLP)}}.
  \bibinfo{pages}{6769--6781}.
\newblock


\bibitem[\protect\citeauthoryear{Kwiatkowski, Palomaki, Redfield, Collins,
  Parikh, Alberti, Epstein, Polosukhin, Devlin, Lee, et~al\mbox{.}}{Kwiatkowski
  et~al\mbox{.}}{2019}]%
        {kwiatkowski2019natural}
\bibfield{author}{\bibinfo{person}{Tom Kwiatkowski},
  \bibinfo{person}{Jennimaria Palomaki}, \bibinfo{person}{Olivia Redfield},
  \bibinfo{person}{Michael Collins}, \bibinfo{person}{Ankur Parikh},
  \bibinfo{person}{Chris Alberti}, \bibinfo{person}{Danielle Epstein},
  \bibinfo{person}{Illia Polosukhin}, \bibinfo{person}{Jacob Devlin},
  \bibinfo{person}{Kenton Lee}, {et~al\mbox{.}}}
  \bibinfo{year}{2019}\natexlab{}.
\newblock \showarticletitle{Natural questions: a benchmark for question
  answering research}.
\newblock \bibinfo{journal}{\emph{Transactions of the Association for
  Computational Linguistics}}  \bibinfo{volume}{7} (\bibinfo{year}{2019}),
  \bibinfo{pages}{453--466}.
\newblock


\bibitem[\protect\citeauthoryear{Ma, Pradeep, Nogueira, and Lin}{Ma
  et~al\mbox{.}}{2022}]%
        {Ma2022DocumentEB}
\bibfield{author}{\bibinfo{person}{Xueguang Ma}, \bibinfo{person}{Ronak
  Pradeep}, \bibinfo{person}{Rodrigo Nogueira}, {and} \bibinfo{person}{Jimmy
  Lin}.} \bibinfo{year}{2022}\natexlab{}.
\newblock \showarticletitle{Document Expansion Baselines and Learned Sparse
  Lexical Representations for MS MARCO V1 and V2}.
\newblock \bibinfo{journal}{\emph{Proceedings of the 45th International ACM
  SIGIR Conference on Research and Development in Information Retrieval}}
  (\bibinfo{year}{2022}).
\newblock


\bibitem[\protect\citeauthoryear{Mehta, Gupta, Tay, Dehghani, Tran, Rao,
  Najork, Strubell, and Metzler}{Mehta et~al\mbox{.}}{2022}]%
        {mehta2022dsi}
\bibfield{author}{\bibinfo{person}{Sanket~Vaibhav Mehta}, \bibinfo{person}{Jai
  Gupta}, \bibinfo{person}{Yi Tay}, \bibinfo{person}{Mostafa Dehghani},
  \bibinfo{person}{Vinh~Q. Tran}, \bibinfo{person}{Jinfeng Rao},
  \bibinfo{person}{Marc Najork}, \bibinfo{person}{Emma Strubell}, {and}
  \bibinfo{person}{Donald Metzler}.} \bibinfo{year}{2022}\natexlab{}.
\newblock \bibinfo{title}{DSI++: Updating Transformer Memory with New
  Documents}.
\newblock
\newblock
\showeprint[arxiv]{2212.09744}~[cs.CL]


\bibitem[\protect\citeauthoryear{Metzler, Tay, Bahri, and Najork}{Metzler
  et~al\mbox{.}}{2021}]%
        {metzler21rethinking}
\bibfield{author}{\bibinfo{person}{Donald Metzler}, \bibinfo{person}{Yi Tay},
  \bibinfo{person}{Dara Bahri}, {and} \bibinfo{person}{Marc Najork}.}
  \bibinfo{year}{2021}\natexlab{}.
\newblock \showarticletitle{Rethinking Search: Making Domain Experts out of
  Dilettantes}.
\newblock \bibinfo{journal}{\emph{SIGIR Forum}} \bibinfo{volume}{55},
  \bibinfo{number}{1}, Article \bibinfo{articleno}{13} (\bibinfo{date}{jul}
  \bibinfo{year}{2021}), \bibinfo{numpages}{27}~pages.
\newblock
\showISSN{0163-5840}
\urldef\tempurl%
\url{https://doi.org/10.1145/3476415.3476428}
\showDOI{\tempurl}


\bibitem[\protect\citeauthoryear{Nguyen, Rosenberg, Song, Gao, Tiwary,
  Majumder, and Deng}{Nguyen et~al\mbox{.}}{2016}]%
        {nguyen2016ms}
\bibfield{author}{\bibinfo{person}{Tri Nguyen}, \bibinfo{person}{Mir
  Rosenberg}, \bibinfo{person}{Xia Song}, \bibinfo{person}{Jianfeng Gao},
  \bibinfo{person}{Saurabh Tiwary}, \bibinfo{person}{Rangan Majumder}, {and}
  \bibinfo{person}{Li Deng}.} \bibinfo{year}{2016}\natexlab{}.
\newblock \showarticletitle{MS MARCO: A human generated machine reading
  comprehension dataset}. In \bibinfo{booktitle}{\emph{CoCo@ NIPS}}.
\newblock


\bibitem[\protect\citeauthoryear{Ni, Abrego, Constant, Ma, Hall, Cer, and
  Yang}{Ni et~al\mbox{.}}{2022a}]%
        {ni22st5}
\bibfield{editor}{\bibinfo{person}{Jianmo Ni},
  \bibinfo{person}{Gustavo~Hernandez Abrego}, \bibinfo{person}{Noah Constant},
  \bibinfo{person}{Ji Ma}, \bibinfo{person}{Keith~B. Hall},
  \bibinfo{person}{Daniel Cer}, {and} \bibinfo{person}{Yinfei Yang}} (Eds.).
  \bibinfo{year}{2022}\natexlab{a}.
\newblock \bibinfo{booktitle}{\emph{Sentence-T5: Scaling up Sentence Encoder
  from Pre-trained Text-to-Text Transfer Transformer}}.
\newblock
\urldef\tempurl%
\url{https://aclanthology.org/2022.findings-acl.146/}
\showURL{%
\tempurl}


\bibitem[\protect\citeauthoryear{Ni, Qu, Lu, Dai, Abrego, Ma, Zhao, Luan, Hall,
  Chang, and Yang}{Ni et~al\mbox{.}}{2022b}]%
        {ni22gtr}
\bibfield{editor}{\bibinfo{person}{Jianmo Ni}, \bibinfo{person}{Chen Qu},
  \bibinfo{person}{Jing Lu}, \bibinfo{person}{Zhuyun Dai},
  \bibinfo{person}{Gustavo~Hernandez Abrego}, \bibinfo{person}{Ji Ma},
  \bibinfo{person}{Vincent Zhao}, \bibinfo{person}{Yi Luan},
  \bibinfo{person}{Keith~B. Hall}, \bibinfo{person}{Ming-Wei Chang}, {and}
  \bibinfo{person}{Yinfei Yang}} (Eds.). \bibinfo{year}{2022}\natexlab{b}.
\newblock \bibinfo{booktitle}{\emph{Large Dual Encoders Are Generalizable
  Retrievers}}.
\newblock
\urldef\tempurl%
\url{https://preview.aclanthology.org/emnlp-22-ingestion/2022.emnlp-main.669.pdf}
\showURL{%
\tempurl}


\bibitem[\protect\citeauthoryear{Nogueira, Jiang, Pradeep, and Lin}{Nogueira
  et~al\mbox{.}}{2020}]%
        {nogueira-etal-2020-document}
\bibfield{author}{\bibinfo{person}{Rodrigo Nogueira}, \bibinfo{person}{Zhiying
  Jiang}, \bibinfo{person}{Ronak Pradeep}, {and} \bibinfo{person}{Jimmy Lin}.}
  \bibinfo{year}{2020}\natexlab{}.
\newblock \showarticletitle{Document Ranking with a Pretrained
  Sequence-to-Sequence Model}. In \bibinfo{booktitle}{\emph{Findings of the
  Association for Computational Linguistics: EMNLP 2020}}.
  \bibinfo{publisher}{Association for Computational Linguistics},
  \bibinfo{address}{Online}, \bibinfo{pages}{708--718}.
\newblock
\urldef\tempurl%
\url{https://doi.org/10.18653/v1/2020.findings-emnlp.63}
\showDOI{\tempurl}


\bibitem[\protect\citeauthoryear{Nogueira, Lin, and Epistemic}{Nogueira
  et~al\mbox{.}}{2019a}]%
        {nogueira2019docT5query}
\bibfield{author}{\bibinfo{person}{Rodrigo Nogueira}, \bibinfo{person}{Jimmy
  Lin}, {and} \bibinfo{person}{AI Epistemic}.}
  \bibinfo{year}{2019}\natexlab{a}.
\newblock \showarticletitle{From doc2query to {docTTTTTquery}}.
\newblock \bibinfo{journal}{\emph{Online preprint}} (\bibinfo{year}{2019}).
\newblock


\bibitem[\protect\citeauthoryear{Nogueira, Yang, Cho, and Lin}{Nogueira
  et~al\mbox{.}}{2019b}]%
        {monobert}
\bibfield{author}{\bibinfo{person}{Rodrigo Nogueira}, \bibinfo{person}{Wei
  Yang}, \bibinfo{person}{Kyunghyun Cho}, {and} \bibinfo{person}{Jimmy Lin}.}
  \bibinfo{year}{2019}\natexlab{b}.
\newblock \showarticletitle{Multi-Stage Document Ranking with {BERT}}.
\newblock \bibinfo{journal}{\emph{CoRR}}  \bibinfo{volume}{abs/1910.14424}
  (\bibinfo{year}{2019}).
\newblock
\showeprint[arXiv]{1910.14424}
\urldef\tempurl%
\url{http://arxiv.org/abs/1910.14424}
\showURL{%
\tempurl}


\bibitem[\protect\citeauthoryear{Nogueira, Yang, Lin, and Cho}{Nogueira
  et~al\mbox{.}}{2019c}]%
        {nogueira2019doc2query}
\bibfield{author}{\bibinfo{person}{Rodrigo Nogueira}, \bibinfo{person}{Wei
  Yang}, \bibinfo{person}{Jimmy Lin}, {and} \bibinfo{person}{Kyunghyun Cho}.}
  \bibinfo{year}{2019}\natexlab{c}.
\newblock \showarticletitle{Document expansion by query prediction}.
\newblock \bibinfo{journal}{\emph{arXiv preprint arXiv:1904.08375}}
  (\bibinfo{year}{2019}).
\newblock


\bibitem[\protect\citeauthoryear{Pradeep, Li, Wang, and Lin}{Pradeep
  et~al\mbox{.}}{2022}]%
        {pradeep22ct}
\bibfield{author}{\bibinfo{person}{Ronak Pradeep}, \bibinfo{person}{Yilin Li},
  \bibinfo{person}{Yuetong Wang}, {and} \bibinfo{person}{Jimmy Lin}.}
  \bibinfo{year}{2022}\natexlab{}.
\newblock \showarticletitle{Neural Query Synthesis and Domain-Specific Ranking
  Templates for Multi-Stage Clinical Trial Matching}. In
  \bibinfo{booktitle}{\emph{Proceedings of the 45th International ACM SIGIR
  Conference on Research and Development in Information Retrieval}} (Madrid,
  Spain) \emph{(\bibinfo{series}{SIGIR '22})}. \bibinfo{publisher}{Association
  for Computing Machinery}, \bibinfo{address}{New York, NY, USA},
  \bibinfo{pages}{2325–2330}.
\newblock
\showISBNx{9781450387323}
\urldef\tempurl%
\url{https://doi.org/10.1145/3477495.3531853}
\showDOI{\tempurl}


\bibitem[\protect\citeauthoryear{Pradeep, Ma, Nogueira, and Lin}{Pradeep
  et~al\mbox{.}}{2021a}]%
        {pradeep21vera}
\bibfield{author}{\bibinfo{person}{Ronak Pradeep}, \bibinfo{person}{Xueguang
  Ma}, \bibinfo{person}{Rodrigo Nogueira}, {and} \bibinfo{person}{Jimmy Lin}.}
  \bibinfo{year}{2021}\natexlab{a}.
\newblock \showarticletitle{Vera: Prediction Techniques for Reducing Harmful
  Misinformation in Consumer Health Search}. In
  \bibinfo{booktitle}{\emph{Proceedings of the 44th International ACM SIGIR
  Conference on Research and Development in Information Retrieval}} (Virtual
  Event, Canada) \emph{(\bibinfo{series}{SIGIR '21})}.
  \bibinfo{publisher}{Association for Computing Machinery},
  \bibinfo{address}{New York, NY, USA}, \bibinfo{pages}{2066–2070}.
\newblock
\showISBNx{9781450380379}
\urldef\tempurl%
\url{https://doi.org/10.1145/3404835.3463120}
\showDOI{\tempurl}


\bibitem[\protect\citeauthoryear{Pradeep, Nogueira, and Lin}{Pradeep
  et~al\mbox{.}}{2021b}]%
        {Pradeep2021TheED}
\bibfield{author}{\bibinfo{person}{Ronak Pradeep}, \bibinfo{person}{Rodrigo
  Nogueira}, {and} \bibinfo{person}{Jimmy~J. Lin}.}
  \bibinfo{year}{2021}\natexlab{b}.
\newblock \showarticletitle{The Expando-Mono-Duo Design Pattern for Text
  Ranking with Pretrained Sequence-to-Sequence Models}.
\newblock \bibinfo{journal}{\emph{ArXiv}}  \bibinfo{volume}{abs/2101.05667}
  (\bibinfo{year}{2021}).
\newblock


\bibitem[\protect\citeauthoryear{Raffel, Shazeer, Roberts, Lee, Narang, Matena,
  Zhou, Li, and Liu}{Raffel et~al\mbox{.}}{2020a}]%
        {raffel2020exploring}
\bibfield{author}{\bibinfo{person}{Colin Raffel}, \bibinfo{person}{Noam
  Shazeer}, \bibinfo{person}{Adam Roberts}, \bibinfo{person}{Katherine Lee},
  \bibinfo{person}{Sharan Narang}, \bibinfo{person}{Michael Matena},
  \bibinfo{person}{Yanqi Zhou}, \bibinfo{person}{Wei Li}, {and}
  \bibinfo{person}{Peter~J Liu}.} \bibinfo{year}{2020}\natexlab{a}.
\newblock \showarticletitle{Exploring the Limits of Transfer Learning with a
  Unified Text-to-Text Transformer}.
\newblock \bibinfo{journal}{\emph{Journal of Machine Learning Research}}
  \bibinfo{volume}{21} (\bibinfo{year}{2020}), \bibinfo{pages}{1--67}.
\newblock


\bibitem[\protect\citeauthoryear{Raffel, Shazeer, Roberts, Lee, Narang, Matena,
  Zhou, Li, and Liu}{Raffel et~al\mbox{.}}{2020b}]%
        {2020t5}
\bibfield{author}{\bibinfo{person}{Colin Raffel}, \bibinfo{person}{Noam
  Shazeer}, \bibinfo{person}{Adam Roberts}, \bibinfo{person}{Katherine Lee},
  \bibinfo{person}{Sharan Narang}, \bibinfo{person}{Michael Matena},
  \bibinfo{person}{Yanqi Zhou}, \bibinfo{person}{Wei Li}, {and}
  \bibinfo{person}{Peter~J. Liu}.} \bibinfo{year}{2020}\natexlab{b}.
\newblock \showarticletitle{Exploring the Limits of Transfer Learning with a
  Unified Text-to-Text Transformer}.
\newblock \bibinfo{journal}{\emph{Journal of Machine Learning Research}}
  \bibinfo{volume}{21}, \bibinfo{number}{140} (\bibinfo{year}{2020}),
  \bibinfo{pages}{1--67}.
\newblock
\urldef\tempurl%
\url{http://jmlr.org/papers/v21/20-074.html}
\showURL{%
\tempurl}


\bibitem[\protect\citeauthoryear{Rajput, Mehta, Singh, Keshavan, Vu, Heldt,
  Hong, Tay, Tran, Samost, Kula, Chi, and Sathiamoorthy}{Rajput
  et~al\mbox{.}}{2023}]%
        {dsi-recsys}
\bibfield{author}{\bibinfo{person}{Shashank Rajput}, \bibinfo{person}{Nikhil
  Mehta}, \bibinfo{person}{Anima Singh}, \bibinfo{person}{Raghunandan~H.
  Keshavan}, \bibinfo{person}{Trung Vu}, \bibinfo{person}{Lukasz Heldt},
  \bibinfo{person}{Lichan Hong}, \bibinfo{person}{Yi Tay},
  \bibinfo{person}{Vinh~Q. Tran}, \bibinfo{person}{Jonah Samost},
  \bibinfo{person}{Maciej Kula}, \bibinfo{person}{Ed~H. Chi}, {and}
  \bibinfo{person}{Maheswaran Sathiamoorthy}.} \bibinfo{year}{2023}\natexlab{}.
\newblock \bibinfo{title}{Recommender Systems with Generative Retrieval}.
\newblock
\newblock
\showeprint[arxiv]{2305.05065}~[cs.IR]


\bibitem[\protect\citeauthoryear{Roberts, Chung, Levskaya, Mishra, Bradbury,
  Andor, Narang, Lester, Gaffney, Mohiuddin, Hawthorne, Lewkowycz, Salcianu,
  van Zee, Austin, Goodman, Soares, Hu, Tsvyashchenko, Chowdhery, Bastings,
  Bulian, Garcia, Ni, Chen, Kenealy, Clark, Lee, Garrette, Lee-Thorp, Raffel,
  Shazeer, Ritter, Bosma, Passos, Maitin-Shepard, Fiedel, Omernick, Saeta,
  Sepassi, Spiridonov, Newlan, and Gesmundo}{Roberts et~al\mbox{.}}{2022}]%
        {roberts2022t5x}
\bibfield{author}{\bibinfo{person}{Adam Roberts}, \bibinfo{person}{Hyung~Won
  Chung}, \bibinfo{person}{Anselm Levskaya}, \bibinfo{person}{Gaurav Mishra},
  \bibinfo{person}{James Bradbury}, \bibinfo{person}{Daniel Andor},
  \bibinfo{person}{Sharan Narang}, \bibinfo{person}{Brian Lester},
  \bibinfo{person}{Colin Gaffney}, \bibinfo{person}{Afroz Mohiuddin},
  \bibinfo{person}{Curtis Hawthorne}, \bibinfo{person}{Aitor Lewkowycz},
  \bibinfo{person}{Alex Salcianu}, \bibinfo{person}{Marc van Zee},
  \bibinfo{person}{Jacob Austin}, \bibinfo{person}{Sebastian Goodman},
  \bibinfo{person}{Livio~Baldini Soares}, \bibinfo{person}{Haitang Hu},
  \bibinfo{person}{Sasha Tsvyashchenko}, \bibinfo{person}{Aakanksha Chowdhery},
  \bibinfo{person}{Jasmijn Bastings}, \bibinfo{person}{Jannis Bulian},
  \bibinfo{person}{Xavier Garcia}, \bibinfo{person}{Jianmo Ni},
  \bibinfo{person}{Andrew Chen}, \bibinfo{person}{Kathleen Kenealy},
  \bibinfo{person}{Jonathan~H. Clark}, \bibinfo{person}{Stephan Lee},
  \bibinfo{person}{Dan Garrette}, \bibinfo{person}{James Lee-Thorp},
  \bibinfo{person}{Colin Raffel}, \bibinfo{person}{Noam Shazeer},
  \bibinfo{person}{Marvin Ritter}, \bibinfo{person}{Maarten Bosma},
  \bibinfo{person}{Alexandre Passos}, \bibinfo{person}{Jeremy Maitin-Shepard},
  \bibinfo{person}{Noah Fiedel}, \bibinfo{person}{Mark Omernick},
  \bibinfo{person}{Brennan Saeta}, \bibinfo{person}{Ryan Sepassi},
  \bibinfo{person}{Alexander Spiridonov}, \bibinfo{person}{Joshua Newlan},
  {and} \bibinfo{person}{Andrea Gesmundo}.} \bibinfo{year}{2022}\natexlab{}.
\newblock \showarticletitle{Scaling Up Models and Data with $\texttt{t5x}$ and
  $\texttt{seqio}$}.
\newblock \bibinfo{journal}{\emph{arXiv preprint arXiv:2203.17189}}
  (\bibinfo{year}{2022}).
\newblock
\urldef\tempurl%
\url{https://arxiv.org/abs/2203.17189}
\showURL{%
\tempurl}


\bibitem[\protect\citeauthoryear{Robertson and Zaragoza}{Robertson and
  Zaragoza}{2009}]%
        {robertson2009probabilistic}
\bibfield{author}{\bibinfo{person}{Stephen Robertson} {and}
  \bibinfo{person}{Hugo Zaragoza}.} \bibinfo{year}{2009}\natexlab{}.
\newblock \bibinfo{booktitle}{\emph{The probabilistic relevance framework: BM25
  and beyond}}.
\newblock \bibinfo{publisher}{Now Publishers Inc}.
\newblock


\bibitem[\protect\citeauthoryear{Sun, Yan, Chen, Wang, Zhu, Ren, Chen, Yin,
  de~Rijke, and Ren}{Sun et~al\mbox{.}}{2023}]%
        {sun2023learning}
\bibfield{author}{\bibinfo{person}{Weiwei Sun}, \bibinfo{person}{Lingyong Yan},
  \bibinfo{person}{Zheng Chen}, \bibinfo{person}{Shuaiqiang Wang},
  \bibinfo{person}{Haichao Zhu}, \bibinfo{person}{Pengjie Ren},
  \bibinfo{person}{Zhumin Chen}, \bibinfo{person}{Dawei Yin},
  \bibinfo{person}{Maarten de Rijke}, {and} \bibinfo{person}{Zhaochun Ren}.}
  \bibinfo{year}{2023}\natexlab{}.
\newblock \bibinfo{title}{Learning to Tokenize for Generative Retrieval}.
\newblock
\newblock
\showeprint[arxiv]{2304.04171}~[cs.IR]


\bibitem[\protect\citeauthoryear{Sutskever, Vinyals, and Le}{Sutskever
  et~al\mbox{.}}{2014}]%
        {sutskever2014sequence}
\bibfield{author}{\bibinfo{person}{Ilya Sutskever}, \bibinfo{person}{Oriol
  Vinyals}, {and} \bibinfo{person}{Quoc~V Le}.}
  \bibinfo{year}{2014}\natexlab{}.
\newblock \showarticletitle{Sequence to Sequence Learning with Neural
  Networks}.
\newblock \bibinfo{journal}{\emph{arXiv preprint arXiv:1409.3215}}
  (\bibinfo{year}{2014}).
\newblock


\bibitem[\protect\citeauthoryear{Tay, Tran, Dehghani, Ni, Bahri, Mehta, Qin,
  Hui, Zhao, Gupta, Schuster, Cohen, and Metzler}{Tay et~al\mbox{.}}{2022}]%
        {Tay2022TransformerMA}
\bibfield{author}{\bibinfo{person}{Yi Tay}, \bibinfo{person}{Vinh~Q. Tran},
  \bibinfo{person}{Mostafa Dehghani}, \bibinfo{person}{Jianmo Ni},
  \bibinfo{person}{Dara Bahri}, \bibinfo{person}{Harsh Mehta},
  \bibinfo{person}{Zhen Qin}, \bibinfo{person}{Kai Hui}, \bibinfo{person}{Zhe
  Zhao}, \bibinfo{person}{Jai Gupta}, \bibinfo{person}{Tal Schuster},
  \bibinfo{person}{William~W. Cohen}, {and} \bibinfo{person}{Donald Metzler}.}
  \bibinfo{year}{2022}\natexlab{}.
\newblock \showarticletitle{Transformer Memory as a Differentiable Search
  Index}.
\newblock \bibinfo{journal}{\emph{ArXiv}}  \bibinfo{volume}{abs/2202.06991}
  (\bibinfo{year}{2022}).
\newblock


\bibitem[\protect\citeauthoryear{Vanderkam, Schonberger, Rowley, and
  Kumar}{Vanderkam et~al\mbox{.}}{2013}]%
        {Vanderkam2013NearestNS}
\bibfield{author}{\bibinfo{person}{Dan Vanderkam}, \bibinfo{person}{Robert~B
  Schonberger}, \bibinfo{person}{H. Rowley}, {and} \bibinfo{person}{Sanjiv
  Kumar}.} \bibinfo{year}{2013}\natexlab{}.
\newblock \showarticletitle{Nearest Neighbor Search in Google Correlate}.
\newblock


\bibitem[\protect\citeauthoryear{Vaswani, Shazeer, Parmar, Uszkoreit, Jones,
  Gomez, Kaiser, and Polosukhin}{Vaswani et~al\mbox{.}}{2017}]%
        {vaswani2017attention}
\bibfield{author}{\bibinfo{person}{Ashish Vaswani}, \bibinfo{person}{Noam
  Shazeer}, \bibinfo{person}{Niki Parmar}, \bibinfo{person}{Jakob Uszkoreit},
  \bibinfo{person}{Llion Jones}, \bibinfo{person}{Aidan~N Gomez},
  \bibinfo{person}{{\L}ukasz Kaiser}, {and} \bibinfo{person}{Illia
  Polosukhin}.} \bibinfo{year}{2017}\natexlab{}.
\newblock \showarticletitle{Attention is all you need}. In
  \bibinfo{booktitle}{\emph{Advances in neural information processing
  systems}}. \bibinfo{pages}{5998--6008}.
\newblock


\bibitem[\protect\citeauthoryear{Wang, Hou, Wang, Miao, Wu, Sun, Chen, Xia,
  Chi, Zhao, Liu, Xie, Sun, Deng, Zhang, and Yang}{Wang et~al\mbox{.}}{2022}]%
        {Wang2022ANC}
\bibfield{author}{\bibinfo{person}{Yujing Wang}, \bibinfo{person}{Ying Hou},
  \bibinfo{person}{Hong Wang}, \bibinfo{person}{Ziming Miao},
  \bibinfo{person}{Shibin Wu}, \bibinfo{person}{Hao Sun}, \bibinfo{person}{Qi
  Chen}, \bibinfo{person}{Yuqing Xia}, \bibinfo{person}{Chengmin Chi},
  \bibinfo{person}{Guoshuai Zhao}, \bibinfo{person}{Zheng Liu},
  \bibinfo{person}{Xing Xie}, \bibinfo{person}{Hao Sun},
  \bibinfo{person}{Weiwei Deng}, \bibinfo{person}{Qi Zhang}, {and}
  \bibinfo{person}{Mao Yang}.} \bibinfo{year}{2022}\natexlab{}.
\newblock \showarticletitle{A Neural Corpus Indexer for Document Retrieval}.
\newblock \bibinfo{journal}{\emph{ArXiv}}  \bibinfo{volume}{abs/2206.02743}
  (\bibinfo{year}{2022}).
\newblock


\bibitem[\protect\citeauthoryear{Zhang, Zhang, Chen, Wang, Chen, Xie, Sun,
  Deng, Zhang, Yang, Yang, Liao, and Guo}{Zhang et~al\mbox{.}}{2023}]%
        {zhang2023irgen}
\bibfield{author}{\bibinfo{person}{Yidan Zhang}, \bibinfo{person}{Ting Zhang},
  \bibinfo{person}{Dong Chen}, \bibinfo{person}{Yujing Wang},
  \bibinfo{person}{Qi Chen}, \bibinfo{person}{Xing Xie}, \bibinfo{person}{Hao
  Sun}, \bibinfo{person}{Weiwei Deng}, \bibinfo{person}{Qi Zhang},
  \bibinfo{person}{Fan Yang}, \bibinfo{person}{Mao Yang},
  \bibinfo{person}{Qingmin Liao}, {and} \bibinfo{person}{Baining Guo}.}
  \bibinfo{year}{2023}\natexlab{}.
\newblock \bibinfo{title}{IRGen: Generative Modeling for Image Retrieval}.
\newblock
\newblock
\showeprint[arxiv]{2303.10126}~[cs.CV]


\bibitem[\protect\citeauthoryear{Zhou, Yao, Dou, Wu, Zhang, and Wen}{Zhou
  et~al\mbox{.}}{2022}]%
        {ultron}
\bibfield{author}{\bibinfo{person}{Yujia Zhou}, \bibinfo{person}{Jing Yao},
  \bibinfo{person}{Zhicheng Dou}, \bibinfo{person}{Ledell Wu},
  \bibinfo{person}{Peitian Zhang}, {and} \bibinfo{person}{Ji-Rong Wen}.}
  \bibinfo{year}{2022}\natexlab{}.
\newblock \bibinfo{title}{Ultron: An Ultimate Retriever on Corpus with a
  Model-based Indexer}.
\newblock
\newblock
\urldef\tempurl%
\url{https://doi.org/10.48550/ARXIV.2208.09257}
\showDOI{\tempurl}


\bibitem[\protect\citeauthoryear{Zhuang, Qin, Jagerman, Hui, Ma, Lu, Ni, Wang,
  and Bendersky}{Zhuang et~al\mbox{.}}{2022a}]%
        {zhuang22rankt5}
\bibfield{author}{\bibinfo{person}{Honglei Zhuang}, \bibinfo{person}{Zhen Qin},
  \bibinfo{person}{Rolf Jagerman}, \bibinfo{person}{Kai Hui},
  \bibinfo{person}{Ji Ma}, \bibinfo{person}{Jing Lu}, \bibinfo{person}{Jianmo
  Ni}, \bibinfo{person}{Xuanhui Wang}, {and} \bibinfo{person}{Michael
  Bendersky}.} \bibinfo{year}{2022}\natexlab{a}.
\newblock \bibinfo{title}{RankT5: Fine-Tuning T5 for Text Ranking with Ranking
  Losses}.
\newblock
\newblock
\urldef\tempurl%
\url{https://doi.org/10.48550/ARXIV.2210.10634}
\showDOI{\tempurl}


\bibitem[\protect\citeauthoryear{Zhuang, Ren, Shou, Pei, Gong, Zuccon, and
  Jiang}{Zhuang et~al\mbox{.}}{2022b}]%
        {zhuang2022bridging}
\bibfield{author}{\bibinfo{person}{Shengyao Zhuang}, \bibinfo{person}{Houxing
  Ren}, \bibinfo{person}{Linjun Shou}, \bibinfo{person}{Jian Pei},
  \bibinfo{person}{Ming Gong}, \bibinfo{person}{Guido Zuccon}, {and}
  \bibinfo{person}{Daxin Jiang}.} \bibinfo{year}{2022}\natexlab{b}.
\newblock \showarticletitle{Bridging the gap between indexing and retrieval for
  differentiable search index with query generation}.
\newblock \bibinfo{journal}{\emph{arXiv preprint arXiv:2206.10128}}
  (\bibinfo{year}{2022}).
\newblock


\bibitem[\protect\citeauthoryear{Ziems, Yu, Zhang, and Jiang}{Ziems
  et~al\mbox{.}}{2023}]%
        {ziems2023large}
\bibfield{author}{\bibinfo{person}{Noah Ziems}, \bibinfo{person}{Wenhao Yu},
  \bibinfo{person}{Zhihan Zhang}, {and} \bibinfo{person}{Meng Jiang}.}
  \bibinfo{year}{2023}\natexlab{}.
\newblock \bibinfo{title}{Large Language Models are Built-in Autoregressive
  Search Engines}.
\newblock
\newblock
\showeprint[arxiv]{2305.09612}~[cs.CL]


\end{thebibliography}

\newpage
%\input{appendix}

% \appendix

\end{document}